\documentclass[epj]{svjour}

\usepackage{bm}
\usepackage{amssymb}
\usepackage{color}
\usepackage{mathtools}
\usepackage{tikz}
\usepackage{url}
\usepackage{hyperref}
\usepackage{cite}
\usepackage{xcolor}
\usepackage{tikz}
\usepackage{amsmath}
\usepackage{float}
 \usepackage{diagbox}
\usepackage{pgfplots}
\usepackage{subcaption}

\usetikzlibrary{positioning}
\usepackage[ruled,vlined]{algorithm2e}

\newcommand{\Eq}[1]{Eq.~\ref{#1}}

\newcommand{\Fig}[1]{Fig.~\ref{#1}}
\newcommand{\App}[1]{App.~\ref{#1}}

\newcommand{\Eqs}[1]{Eqs.~\ref{#1}}

\usepackage{siunitx}
\newcommand{\gev}[1]{\SI{#1}{\giga\electronvolt}}

\graphicspath{{./images/}}

\begin{document}

\title{Compressing PDF sets using generative adversarial networks}

\author{Stefano Carrazza\inst{1,2,3}, Juan Cruz-Martinez\inst{1},  Tanjona R. Rabemananjara\inst{1}}
%
%
\institute{TIF Lab, Dipartimento di Fisica, Universit\`a degli Studi di Milano
and INFN Sezione di Milano. \and CERN,
Theoretical Physics Department, CH-1211 Geneva 23, Switzerland. \and Quantum
Research Centre, Technology Innovation Institute, Abu Dhabi, UAE.}

\date{Received: date / Revised version: date}
\abstract{We present a compression algorithm for parton densities
using synthetic replicas generated from the training of a Generative Adversarial
Network (GAN). The generated replicas are used to further enhance the
statistics of a given Monte Carlo PDF set prior to compression. This results in
a compression methodology that is able to provide a compressed set with smaller
number of replicas and a more adequate representation of the original
probability distribution. We also address the question of whether the GAN could
be used as an alternative mechanism to avoid the fitting of large number of replicas.
} 

\authorrunning{S.C, J.C.M, T.R.R.}

\maketitle
%

\section{Introduction}

Parton distribution functions (PDFs) are crucial ingredients for all predictions of physical
observables at hadron colliders such as the LHC, and efforts to push their uncertainties to smaller
values are becoming increasingly relevant. As a matter of fact, they are one of the
dominant sources of uncertainties in precision measurements. An example of this is
the limitation that PDFs play in the extraction of the Higgs couplings from data~\cite{Cepeda:2019klc}.

In the past few years, several methods have been developed to make the determination of PDFs
more precise. To date, there exists various PDF fitter groups~\cite{Ball:2017nwa, Bailey:2020ooq,
Hou:2019jgw, Alekhin:2013nda} implementing different methodologies and providing PDF sets with
different estimates of the associated uncertainties.
In the NNPDF approach, where the PDFs are represented in terms of an ensemble of
Monte Carlo replicas~\cite{Ball:2017nwa}, a large number of those replicas are
required in order to reproduce the most accurate representation of the underlying probability
distribution. There has been considerable evidence~\cite{Carrazza:2015hva, Badger:2016bpw,
Butterworth:2015oua} that the convergence of the Monte Carlo PDF replicas to the asymptotic result
is slow, and one might require $\mathcal{O}(1000)$ replicas to get accurate results. As a matter of
fact, one of the main differences between a fit with $100$ and $1000$ Monte Carlo replicas is that
correlations between PDFs are reproduced more accurately in the latter~\cite{Carrazza:2015hva}. From a
practical point of view, however, there are challenges related to the use of such a large set of PDF
members. Indeed, having to deal with a large ensemble of replicas when producing phenomenological studies is
not ideal. To address this issue, a compression methodology that reduces the original Monte Carlo PDF
set into a smaller subset was introduced in Ref.~\cite{Carrazza:2015hva}. Conceptually, the compression works
by searching for the subset of replicas that reproduces best the statistical
features of the original prior PDF distribution.

In this paper, we propose a new compression strategy that aims to provide a
compressed set with an even smaller number of replicas when compared to Ref.~\cite{Carrazza:2015hva},
while maintaining an accurate representation of the
original probability distribution. Our approach relies on the deep learning techniques usually
known as Generative Adversarial Neural Networks~\cite{NIPS2014_5ca3e9b1} or GANs in short. GANs belong to the class of
unsupervised machine learning algorithms where a generative model is trained to produce new data
which is indistinguishable under certain criteria from the training data. GANs are mainly used for image
modeling~\cite{zhu2020unpaired}, but in the last couple years, many applications have been found in
High Energy Physics (HEP)~\cite{deOliveira:2017pjk, Paganini:2017dwg, Butter:2019eyo, Bellagente:2019uyp,
Butter:2019cae, Carrazza:2019cnt, Backes:2020vka, Butter:2020tvl, Butter:2020qhk, Matchev:2020tbw}. Here, we propose
to use the GANs to enhance the statistics of a given input Monte Carlo PDF by generating
what we call \emph{synthetic} replicas. This is supported by the following observation:
large replica samples contain fluctuations that average out to the asymptotic limit. In the
standard approach, the job of the PDF compressor is to only extract samples that present small
fluctuations and which reproduce best the statistical properties of the original distribution. It should be
therefore possible to use GANs to generate samples of replicas that contain less fluctuations
and once combined with samples from the prior lead to a more efficient compressed representation
of the full result.

Despite the fact that the techniques described in this paper might be generalizable to produce
larger PDF sets, we emphasize that our main goal is to provide a technique for minimizing the information
loss due to the compression of larger into smaller sets.

The paper is organized as follows. Section~\ref{sec:compressor}
and~\ref{sec:ganpdfs} provide a brief description of the compression and GAN
methodologies respectively. The framework in which the two methodologies are
combined together is described in Section~\ref{sec:gan-compressor}.
Section~\ref{sec:results} presents the results, highlighting the improvement
with respect to the previous compression methodology. Section~\ref{sec:outlook}
gives an outlook of the potential usage of GANs to by-pass the fitting
procedure.

\section{Compression: methodological review}
\label{sec:compressor}

Let us begin by giving a brief review of the compression methodology formally
introduced in Ref.~\cite{Carrazza:2015hva}. The underlying idea behind the compression of (combined)
Monte Carlo PDF replicas consists in finding a subset of the original set of PDF replicas
such that the statistical distance between the original and compressed probability
distributions is minimal.

The compression strategy relies on two main ingredients: first, a proper definition of the
distance metric that measure the difference between the prior and the compressed distributions;
second, an appropriate minimization algorithm that explores the space of minima.

As originally proposed~\cite{Carrazza:2015hva}, a suitable figure of merit to quantify
the distinguishability between the prior and compressed probability distributions is the error function:
\begin{align}
    \mathrm{ERF} = \frac{1}{N_{\text{EST}}} \sum_{k} \frac{1}{N_k} \sum_{i}
    \left( \frac{C^{k}(x_{i}) - P^{k}(x_{i})}{P^{k}(x_{i})} \right)^2 ,
    \label{eq:ERF}
\end{align}
where $N_{\text{EST}}$ denotes the total number of statistical estimators used to quantify the
distance between the prior and compressed PDF sets, $k$ runs over all statistical
estimators with $N_k$ the appropriate normalization factor, $P^{k}(x_i)$ is the value of the
estimator $k$ computed at a given point $i$ in the $x$ grid, and $C^{k}(x_i)$
is the corresponding value of the same estimator for the compressed distribution.
The scale at which the PDFs are computed is fixed and the same for all estimators.
The list of statistical estimators entering the expression of the total error function in
\Eq{eq:ERF} includes lower moments (such as mean and standard deviation) and standardized moments (such
as Skewness and Kurtosis). In addition, in order to preserve higher moments and PDF-induced
correlations in physical cross sections, the Kolmogorov-Smirnov and the correlation between
multiple PDF flavours are also considered. For each estimator, a proper normalization factor
has to be included in order to compensate for the various orders of magnitude in different
regions in the $(x, Q)$ space mainly present in higher moments. For an ample description of the individual
statistical estimators with the respective expression of their normalization factors, we refer
the reader to the original compression paper~\cite{Carrazza:2015hva}.

Once the set of statistical estimators and the target size of the compressed PDF set are defined, the
compression algorithm searches for the combination of replicas that leads to the minimal value of
the total error function. Due to the discrete nature of the compression problem, it is adequate
to perform the minimization using Evolution Algorithm (EA) strategies such as the Genetic
Algorithm (GA) or the Covariance Matrix Adaptation (CMA)
to select the replicas entering the compressed set.

The methodology presented above is currently implemented in a C++ code referred to as
\texttt{compressor}~\cite{Carrazza:2015hva} which uses a GA as the minimization strategy.
In order to construct a new
framework that allows for various methodological enhancements, we have developed a new
compression code  written in python and based on the object oriented approach for grater
flexibility and maintainability.
Henceforth, we refer to the new implementation as \texttt{pyCompressor}~\cite{Compressor:2020}.
The new framework provides additional
features such as the Covariance Matrix Adaptative (CMA) minimization strategy and the
adiabatic minimization procedure, whose relevance will be explained in the next sections.
In addition to the above-mentioned advantages, the new code is also faster. A benchmark
comparison with the \texttt{compressor} is presented in~\App{app:benchmark}.

By default, the \texttt{pyCompressor} computes the input PDF grid for $n_f = 8$ light
partons $(g, u, s, d, c, \bar{u}, \bar{s}, \bar{d})$ at some energy scale
(ex: $Q_0 = \gev{1.65}$).
The range of $x$ points is restricted within the regions where experimental data are
available, namely $(10^{-5}, 1)$. The default estimators are the same as the ones
considered in Ref.~\cite{Carrazza:2015hva}.

\section{How to GAN PDFs?}
\label{sec:ganpdfs}

The following section describes how techniques from generative adversarial models
could be used to improve the efficiency of the compression algorithm. The framework
presented here has been implemented in a standalone python package that we dubbed
\texttt{ganpdfs}~\cite{ganpdfs:2020}. The idea is to use generative neural networks to enhance the
statistics of the original Monte Carlo PDF replicas prior to the compression by
generating synthetic PDF replicas.

\subsection{Introduction to GANs for PDFs}

The problem we are concerned with is the following: suppose our Monte Carlo PDF replicas
follow a probability distribution $p_R$, we would like to generate synthetic
replicas following some probability distribution $p_\theta$ such that $p_\theta$
is very close to $p_R$, i.e, the Kullback-Leibler divergence
\begin{align}
	\mathrm{KL} \left( p_R || p_\theta \right) = \int_x \mathrm{d}x~p_R (x) \log \frac{p_R(x)}{p_\theta(x)}
	\label{eq:kl-div}
\end{align}
is minimal. One way to achieve this is by defining a latent variable $z$ with fixed
probability distribution $p_z$.
$z$ is then passed as the input of a neural network function
$G_\theta : z \to x_g$ that generates samples following $p_\theta$. Hence, by optimizing
the parameters $\theta$ we can modify the distribution $p_\theta$ to approach $p_R$. The
most prominent examples of such a procedure are known as Generative Adversarial Networks
(GANs)~\cite{goodfellow2014generative, zhu2020unpaired, arjovsky2017wasserstein, brock2019large,
radford2016unsupervised}.

Generative adversarial models
involve two main agents known as \emph{generator} and \emph{discriminator}. The generator
$G_\theta$ is a differentiable function represented in our case by a multilayer perceptron whose job is
to deterministically generate samples $x_g$ from the latent variable $z$. The discriminator
$D_\phi$ is also a multilayer perceptron whose main job is to distinguish samples from real
and synthetic PDF replicas.

The generator $G_\theta$ and discriminator $D_\phi$ are then trained in an adversarial
way: $G_\theta$ tries to capture the probability distribution of the input PDF replicas
and generates new samples following the same probability distribution (therefore minimizes
the objective $p_R = p_\theta$), and $D_\phi$ tries to distinguish whether the sample came from
an input PDF rather than from the generator (therefore maximizes the objective $p_R \neq p_\theta$).
This adversarial training allows both models to improve to the point where
the generator is able to create synthetic PDFs such that the discriminator
can no longer distinguish between synthetic and original replicas. This is a sort of min-max game
where the GAN objective function is given by following expression~\cite{goodfellow2014generative}:
\allowdisplaybreaks
\begin{align}
	\min _{\theta} \: \max _{\phi} \: V \left(G_{\theta}, D_{\phi}\right) =& \,
	\mathrm{E}_{x \sim p_R} \left( \log D_\phi (x) \right)
	\nonumber \\
	+ \mathrm{E}_{z \sim p_z} & \left( \log\left( 1 - D_\phi \left( G_\theta(z) \right) \right) \right) .
	\label{eq:objective-function}
\end{align}
For a fixed generator $G_\theta$, the discriminator performs a binary classification by
maximizing $V$ w.r.t. $\phi$ while assigning the probability 1 to samples from the prior
PDF replicas $(x \sim p_R)$, and assigning probability 0 to the synthetic samples
$(x \sim p_\theta)$. Therefore, the discriminator is at its optimal efficiency when:
\begin{align}
    D^{*}(x)=\frac{p_R (x)}{p_R (x) + p_\theta (x)} .
    \label{eq:disc-optimal}
\end{align}
If we now assume that the discriminator is at its best ($D(x) = D^{*}(x)$), then the
objective function that the generator is trying to minimize can be expressed in terms
of the Jensen-Shannon divergence $\mathrm{JSD} (p_R, p_\theta)$ as follows:
\begin{align}
    V \left(G_{\theta}, D^{*}_{\phi}\right) = 2 \, \mathrm{JSD} (p_R, p_\theta) - 2 \log 2,
    \label{eq:gen-optimal}
\end{align}
where the Jensen-Shannon divergence satisfies all the properties of the Kullback-Leibler
divergence and has the additional constraint that $\mathrm{JSD} (p_R, p_\theta) = \mathrm{JSD} (p_\theta, p_R)$.
We then see from \Eqs{eq:objective-function}~-~\ref{eq:gen-optimal} that the best
objective value we can achieve with optimal generator and discriminator is $(- 2 \log 2)$.

The generator and discriminator are trained in the following way: the generator generates
a batch of synthetic PDF samples that along with the samples from the input PDF are
provided to the discriminator \footnote{It is important that during the training of the generator,
the discriminator is not training. As it will be explained later, having an over-optimized
discriminator leads to instabilities.} (see Algorithm \ref{algo:vanialla-gan}).
\begin{algorithm}
\SetAlgoLined
\For{epochs $1, \cdots, N$}{
\For{discriminator steps $1, \cdots, k$}{
	- Sample minibatch of size $m$ from the real input sample:
	$\lbrace x_r^{(1)}, \cdots, x_r^{(m)} \rbrace$ \\
	- Sample minibatch of size $m$ from the latent space:
	$\lbrace z^{(1)}, \cdots, z^{(m)} \rbrace$ \\
	- Perform gradient \textbf{ascent} on discriminator:
	\begin{align*}
	\hspace*{-0.4cm}
	\nabla_{\phi} V\left(G_{\theta}, D_{\phi}\right) &= \frac{1}{m} \nabla_{\phi}
	\sum_{i=1}^{m} \log D_{\phi} \left(x_r^{(i)}\right) \\
    & + \frac{1}{m} \nabla_{\phi} \sum_{i=1}^{m} \log \left(1-D_{\phi}\left(G_{\theta}\left(z^{(i)}\right
	)\right)\right)
	\end{align*}
}
\For{generator steps $1, \cdots, l$}{
	- Sample minibatch of size $m$ from the latent space: $\lbrace z^{(1)}, \cdots,
	z^{(m)} \rbrace$ \\
	- Perform gradient \textbf{descent} on generator:
	\begin{align*}
	\hspace*{-0.4cm}
	\nabla_{\theta} V\left(G_{\theta}, D_{\phi}\right)=\frac{1}{m} \nabla_{\theta}
	\sum_{i=1}^{m} \log \left(1-D_{\phi}\left(G_{\theta}\left(z^{(i)}\right)\right)
	\right)
	\end{align*}
}
}
\caption{Stochastic gradient descent training of generative adversarial networks~\cite{goodfellow2014generative}.}
\label{algo:vanialla-gan}
\end{algorithm}
Algorithm \ref{algo:vanialla-gan} is a typical formulation of a
generative-based adversarial strategy. However, as we will discuss in the next section,
working with such an implementation in practice is very challenging and often leads to
poor results.

\subsection{Challenges in training GANs}

Training generative adversarial models can be very challenging due to the lack of
stopping criteria that estimates when exactly GANs have finished training. Therefore,
there is no guarantee that the equilibrium is reached.
In addition, GANs have common failure modes due to inappropriate choices of
network architecture, loss function or optimization algorithm.
Several solutions have been proposed to address
these issues which is a topic still subject to active research. For
a review, we refer the reader to Refs.~\cite{saxena2020generative, wiatrak2020stabilizing, 8667290}.

In this section, we describe our solutions to some of these challenges
which are specific to our problem is described in the next section.
The most encountered limitations of
GANs in our case are: non-convergence, vanishing gradients, and mode collapse.

It is often the case that during optimization, the losses of generator and
discriminator continue to oscillate without converging to a clear stopping
value. Although the existence of an equilibrium has been proved in the original
GAN paper~\cite{goodfellow2014generative}, there is no guarantee that such an
equilibrium will be reached in practice. This is mainly due to the fact that the
generator and discriminator are modelled in terms of neural networks, thus restricting
the optimization procedures to the parameter space of the networks rather than learning
directly from the probability distributions~\cite{wiatrak2020stabilizing}. On the other
hand, this non-convergence could happen even if the models has ventured near the equilibrium
point. When the generated samples are far from the target distribution, the discriminator
pushes the generator  towards the true data distribution while at the same time increasing
its slope. When the generated samples approaches the target distribution, the discriminator's
slope is the highest, pushing away the generator from the true data
distribution~\cite{mescheder2018training}. As a result, depending on the stopping criteria,
the GAN optimization might not necessarily converge to a Nash equilibrium~\cite{goodfellow2014generative}
in which the performance of the discriminator and generator cannot be improved further
(i.e. optimal conditions). In such a scenario, the generated samples are just collections of random noise.

The vanishing gradients occur when the discriminator is over-optimized and does
not provide enough information to the generator to make substantial progress~\cite{saxena2020generative}.
During backpropagation, the gradient of the generator flows backward from the last
layer to the first, getting smaller at each iteration. This raises some complications since,
in practice, the objective function of a discriminator close to optimal is not
$2 \, \mathrm{JSD} (p_R, p_\theta) - 2\log 2$ as stated in \Eq{eq:gen-optimal}, but rather
close to 0. This pushes the loss function to 0, providing little or no feedback
to the generator. In such a case, the gradient does not change the values of the
weights in the initial layers, altering the training of the subsequent layers.

Finally, the mode collapse occurs when the generator outputs samples of low diversity~\cite{bang2018mggan}, i.e,
the generator maps multiple distinct input to the same output. In our case, this
translates into the GAN generating PDF replicas that capture only a few of the modes
of the input PDF. Mode collapse is a common pathology to all generative adversarial
models as the cause is deeply rooted in the concept of GANs.
\begin{figure}[ht]
    \centering
    \includegraphics[width=\linewidth]{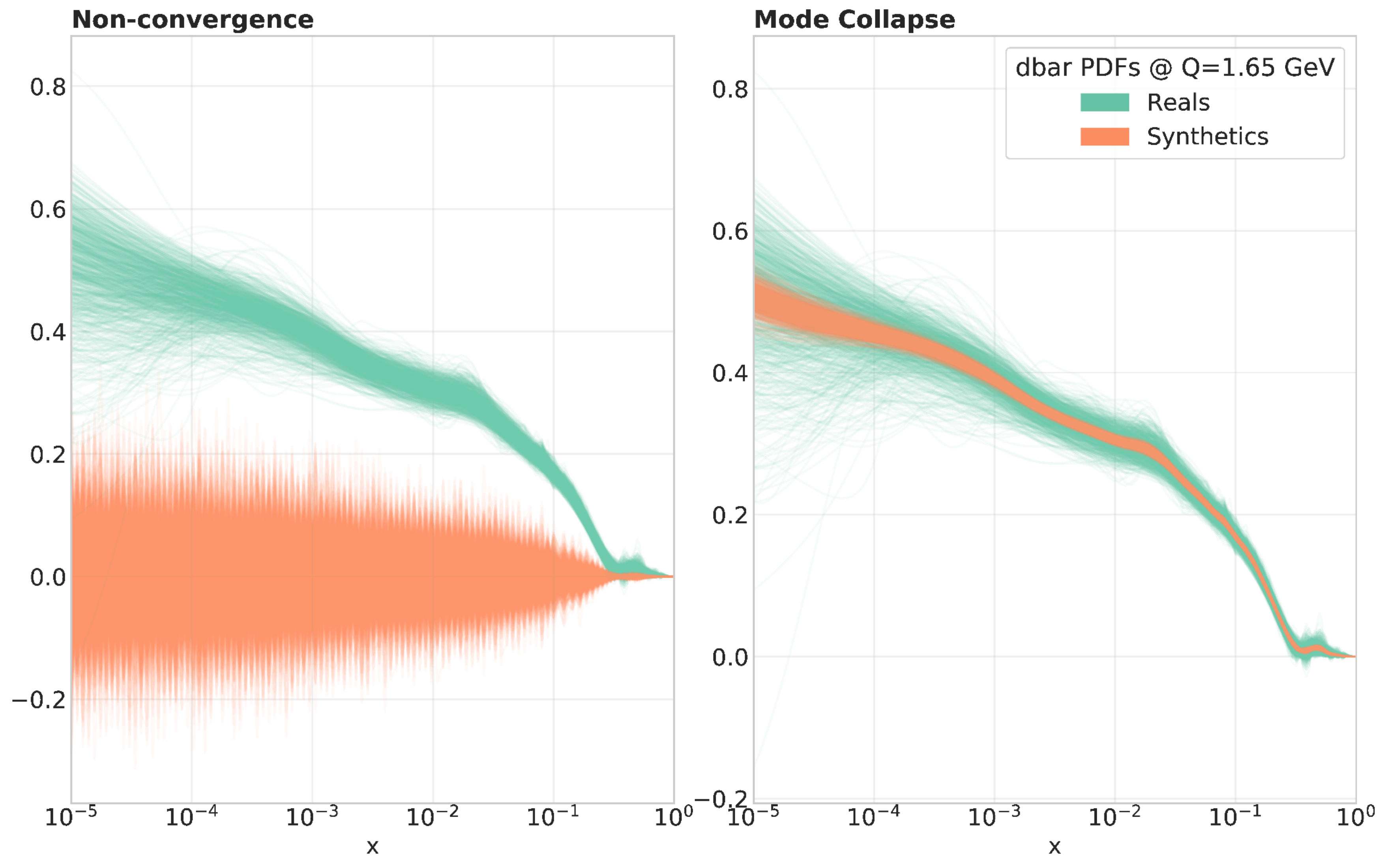}
    \caption{Illustration of non-convergence and mode collapse in the context of Monte Carlo
        PDF replica generation. The two plots were generated using Algorithm~\ref{algo:vanialla-gan} but
    with different neural network architecture.}
\end{figure}

The challenges mentioned above contribute to the instabilities that arise in
training GAN models. What often complicates the situation is that these obstacles
are linked with each other and one might require a procedure that tackles all of them
at the same time. As we will briefly describe in the next section, there exist
various regularization techniques, some specific to the problem in question, that
could influence the stability during the training.

\subsection{The \texttt{ganpdfs} methodology}

In this section, we describe some regularization procedures implemented in
\texttt{ganpdfs} that alleviate the challenges mentioned in the previous
section when using the standard GAN to generate Monte Carlo PDF replicas.

One of the main causes leading to unstable optimization, and therefore to
non-convergence, is the low dimensional support of generated and target distributions~\cite{wiatrak2020stabilizing}.
This could be handled by providing additional information to the input latent
vector $z$~\cite{mescheder2018training}.
This method supplies the latent variable with relevant features
of the real sample. In our case, this is done by taking a combination of
the input PDF replicas, adding on top of it Gaussian noise, and using it as
the latent variable. This has been shown to improve significantly the stability
of the GAN during optimization.

On the other hand, it was shown in Refs.~\cite{mescheder2018adversarial, arjovsky2017principled}
that the objective function that the standard GAN minimizes is not continuous
w.r.t. the generator's parameters. This can lead to both vanishing gradients and mode
collapse problems. Such a shortcoming was already noticed in the original GAN paper
where it was shown that the Jensen-Shannon divergence under idealized conditions
contributes to oscillating behaviours. As a result, a large number of studies have
been devoted to finding a well defined objective function. In our implementation,
we resorted to the Wasserstein or Earth's Mover's (EM) distance~\cite{pinetz2019estimation}
which is implemented in Wasserstein GAN (WGAN~\cite{arjovsky2017wasserstein, gulrajani2017improved}).
The EM loss function is defined in the following way:
\begin{align}
    \min _{\theta} \: \max _{\phi} \, V  = \mathrm{E}_{x \sim p_R} \left( D_\phi (x) \right)
    - \mathrm{E} _{x \sim p_\theta} \left( D_\phi (x_g) \right),
\end{align}
where again $x_g = G_\theta (z), \, z \sim p_z$. The EM distance metric is
effective in solving vanishing gradients and mode collapse as it is continuously
differentiable w.r.t. the generator and discriminator's parameters. As a result,
WGAN models result in a discriminator function whose gradients w.r.t. its input
is better behaved that the standard GAN. This means that discriminator can
be trained until optimality without worrying about vanishing gradients. The default GAN
architecture in \texttt{ganpdfs} is based on WGAN
and is described in Algorithm \ref{algo:wgan}.
\begin{algorithm}
    \SetAlgoLined
    \For{epochs $1, \cdots, N$}{
        \For{discriminator steps $1, \cdots, k$}{
            - Sample minibatch of size $m$ from the real input sample:
            $\lbrace x_r^{(1)}, \cdots, x_r^{(m)} \rbrace$ \\
            - Sample minibatch of size $m$ from the \textbf{custom} latent space:
            $\lbrace z^{(1)}, \cdots, z^{(m)} \rbrace$ \\
            - Perform gradient \textbf{ascent} on discriminator:
            \begin{align*}
                \hspace*{-0.4cm}
                \nabla_{\phi} V\left(G_{\theta}, D_{\phi}\right) = \frac{1}{m} \nabla_{\phi}
                \sum_{i=1}^{m} \left( D_{\phi} \left(x_r^{(i)}\right) - D_\phi \left(x_g^{(i)}
                \right) \right)
            \end{align*} \\
            - $\phi = \phi + \mathrm{RMSProp}(\phi, \nabla_{\phi} V) $ \\
            - Clip weights within $\left[ -c, c \right]$
        }
        \For{generator steps $1, \cdots, l$}{
            - Sample minibatch of size $m$ from the \textbf{custom} latent space:
            $\lbrace z^{(1)}, \cdots, z^{(m)} \rbrace$ \\
            - Perform gradient \textbf{descent} on generator:
            \begin{align*}
                \hspace*{-0.4cm}
                \nabla_{\theta} V\left(G_{\theta}, D_{\phi}\right)=\frac{1}{m} \nabla_{\theta}
                \sum_{i=1}^{m} D_\phi \left( x_g^{(i)} \right)
            \end{align*} \\
            - $\theta = \theta - \mathrm{RMSProp}(\theta, \nabla_{\theta} V) $
        }
    }
    \caption{Stochastic gradient descent training of the GANs implemented in \texttt{ganpdfs}.}
    \label{algo:wgan}
\end{algorithm}

Although the EM distance measure yields non-zero gradients everywhere for the discriminator,
the resulting architecture can still be unstable when gradients of the loss function are large.
This is addressed by clipping the weights of the discriminator to lie within a
compact space defined by $\left[ -c, c \right]$.

Finally, one of the factors that influence the training of GANs is the
architecture of the neural networks. The choice of architecture could scale up
significantly the GAN's performance. However, coming up with values for hyper-parameters
such as the size of the network or the number of nodes in a given layer is particularly
challenging as the parameter space is very large. A heuristic approach
designed to tackle this problem is called hyper-parameter scan or hyper-parameter
optimization. Such an optimization allows for a search of the best
hyper-parameters through an iterative scan of the parameter space. In our
implementation, we rely on the \emph{Tree-structured Parzen Estimator}
(TPE)~\cite{bergstra:hal-00642998} as an optimization algorithm, and we use the Fr\'{e}chet
Inception Distance (FID)~\cite{heusel2018gans} as the figure of merit to hyper-optimize
on. For a target distribution with mean $\mu_r$ and covariance $\Sigma_r$ and a synthetic
distribution with mean $\mu_s$ and covariance $\Sigma_s$, the FID is defined as:
\begin{align}
	\text{FID} = \vert\vert \mu_r - \mu_s \vert\vert^2 + \text{Tr}\left( \Sigma_r + \Sigma_s
	- 2 \sqrt{\Sigma_r \Sigma_s} \right).
	\label{eq:fid}
\end{align}
The smaller the value of the FID is, the closer the generated samples are to the target
distribution. The TPE algorithm will search for sets of hyper-parameter which lead to lower 
values of the FID using the \texttt{hyperopt}~\cite{pmlr-v28-bergstra13} hyper-parameter 
optimization tool.

\begin{figure}[tb]
	\centering
	\includegraphics[width=.95\linewidth]{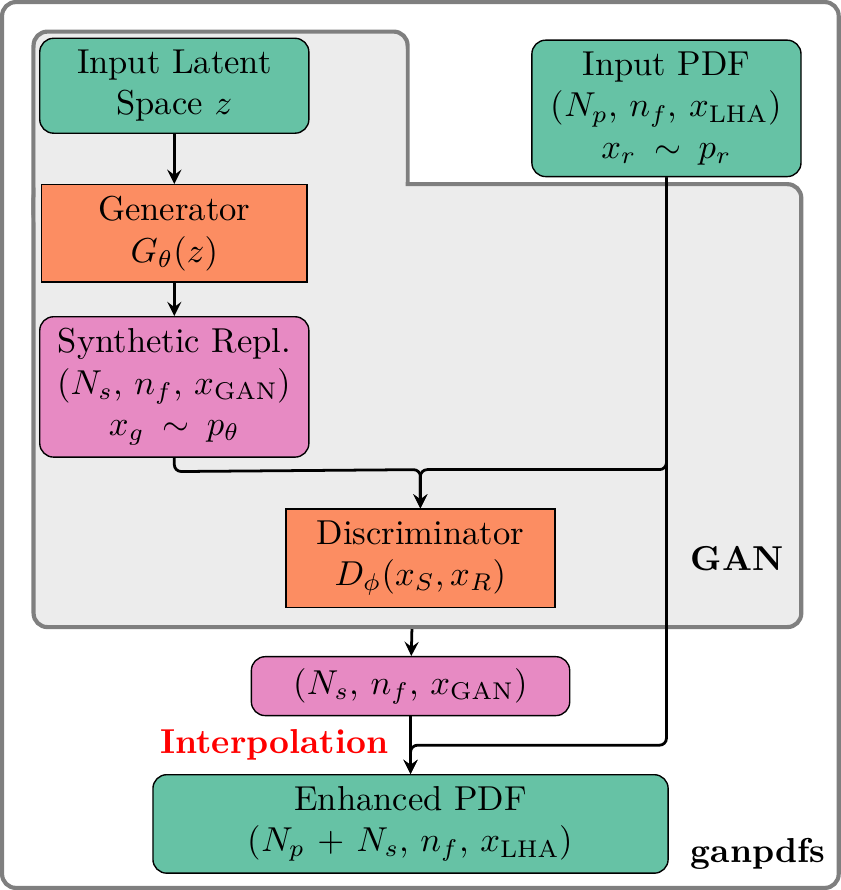}
	\caption{Flowchart describing the combined GANs and compression framework. The
		\texttt{ganpdfs} code is interfaced with the \texttt{pyCompressor} code from which input
		arguments related to the enhancement and compression are passed.}
	\label{fig:gan-standalone}
\end{figure}

The implementation of the GAN for PDFs has been done using the machine learning framework
\texttt{TensorFlow}\cite{tensorflow2015:whitepaper}. A diagrammatic summary of the
workflow is given in~\Fig{fig:gan-standalone}.
We see that in order to train, the GAN receives as input a tensor
with shape ($N_p$, $n_f$, $x_{\text{LHA}}$), where $N_p$ denotes the number of input
replicas, $n_f$ the total number of flavours, and $x_{\text{LHA}}$ the size of the grid in x.

The output of \texttt{ganpdfs} is a LHAPDF grid~\cite{Buckley:2014ana} at a starting scale $Q_0=\gev{1.65}$
which can then be evolved using APFEL~\cite{Bertone:2013vaa} to generate a full LHAPDF set of $N_s$
synthetic replicas. Physical constrains, such as the positivity of the PDFs or the normalization of
the different flavours, are not enforced but rather inferred from the underlying distribution.

\section{The GAN-enhanced compressor}
\label{sec:gan-compressor}

\begin{figure}[tb]
    \centering
    \includegraphics[width=\linewidth]{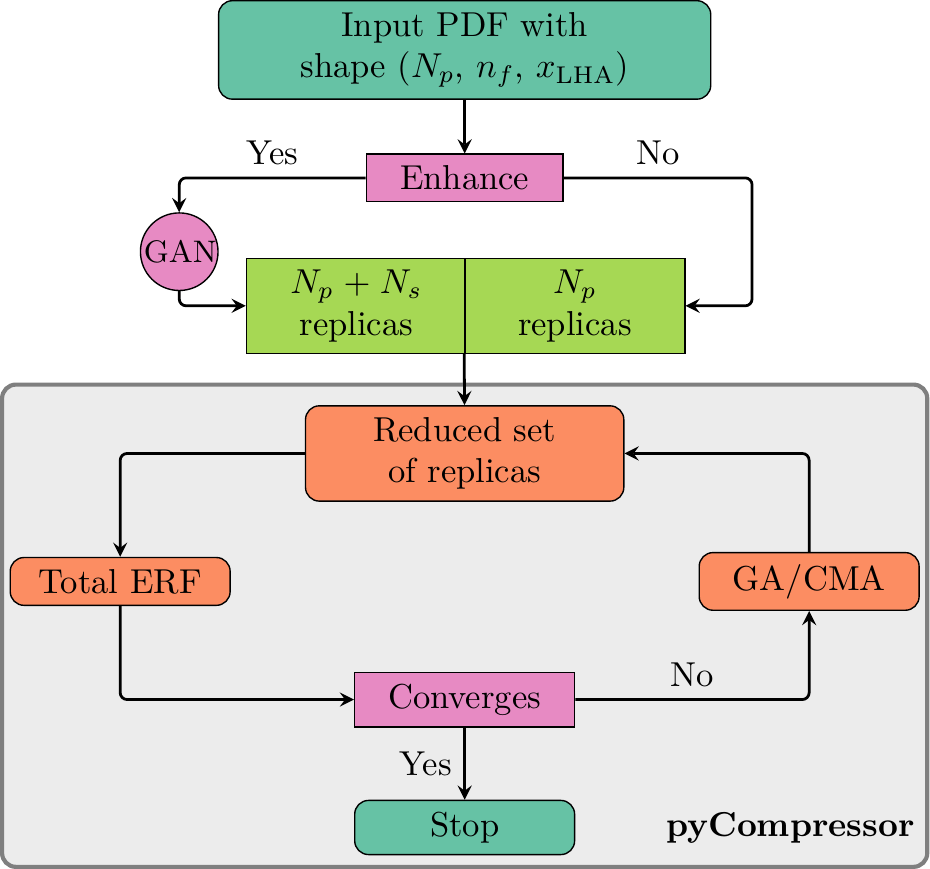}
    \caption{Flowchart describing the combined GANs and compression framework. The
        \texttt{ganpdfs} code is interfaced with the \texttt{pyCompressor} code from which input
    arguments related to the enhancement and compression are passed.}
    \label{fig:full-framework}
\end{figure}

Having introduced the concepts of PDF compression and generative adversarial techniques, we
present in~\Fig{fig:full-framework} a schematic diagram combining the two frameworks. The workflow
goes as follows: the input PDF grid is computed for a given Monte Carlo PDF set containing $N_p$ replicas
at fixed $Q_0$ and at some value of the Bjorken $x$. If GAN enhancement is not required, the reduction
strategy follows the standard compression introduced in Section~\ref{sec:compressor}. If, on the
other hand, the enhancement is required, the GAN is used to generate $N_{s}$ synthetic replicas.
Notice that the format of the $x$ grid in which the GAN is trained does not have to be the same as the
LHAPDF. By default, the $x_{\text{GAN}}$ is a grid of $N_x=500$ points logarithmically spaced in the
small-$x$ region $[10^{-9},10^{-1}]$ and linearly spaced in the large-$x$ region $[10^{-1},1]$. In such
a scenario, an interpolation is required in order to represent the output of the GAN in the LHAPDF format.
The synthetic replicas along with the prior (\emph{enhanced} set henceforth)
has a total size $N_e=N_p+N_s$. The combined sets then passed to the \texttt{pyCompressor} code for
compression. In the context of GAN-enhanced compression, the samples of replicas that will end up
in the compressed set are drawn from the enhanced set rather than from the prior. However, since
we are still trying to reproduce the probability distribution of the original PDF set, the
minimization has to be performed w.r.t. the input Monte Carlo replicas. As a consequence, the
expression of the error function in~\Eq{eq:ERF} has to be modified accordingly, i.e. computing the
estimator $C$ using samples from the enhanced distribution. It is important to emphasize that the
expression of the normalization factors does not change; that is, the random
set of replicas have to be extracted from the prior.

Performing a compression from an enhanced set, however, can be very challenging. Indeed, the
factorial growth of the number of replicas probes the limit of the Genetic Algorithm and therefore
spoils the minimization procedure. In order to address this combinatorial problem, we implemented
in the compression code an adiabatic minimization procedure (see~\App{app:adiabatic} for details).
\begin{figure*}[tb]
    \centering
    \includegraphics[width=\linewidth]{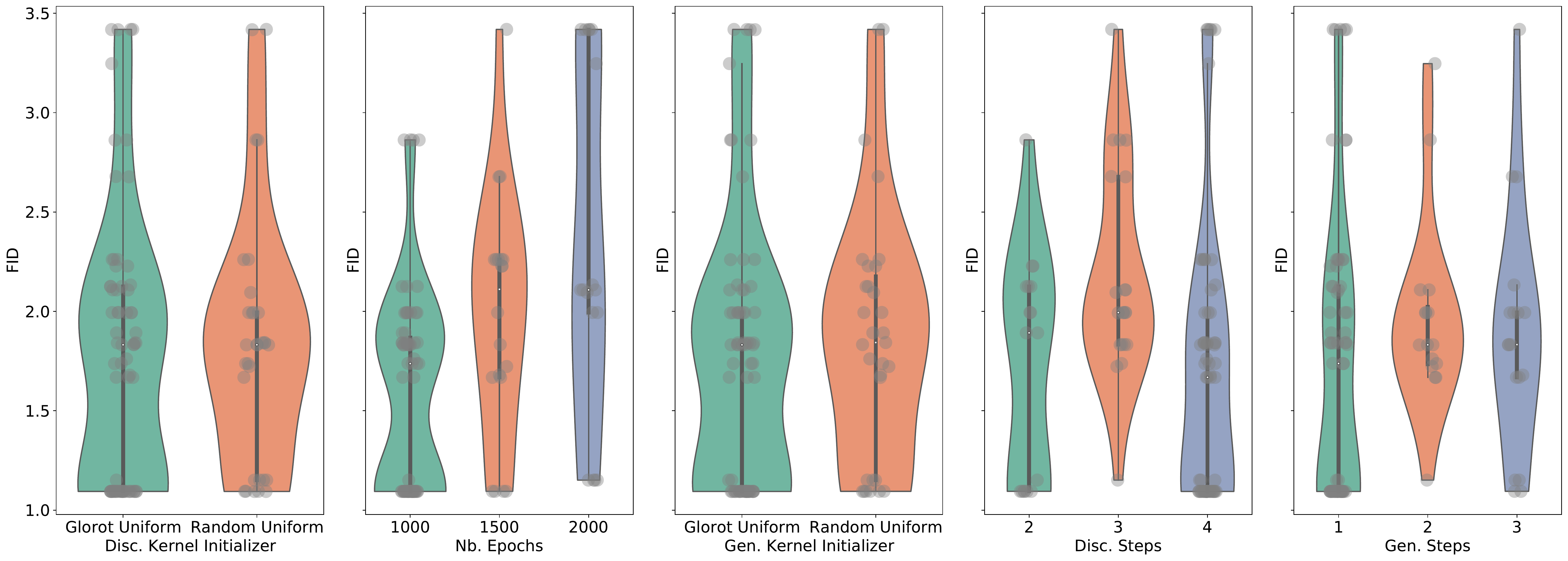}
    \caption{Graphical representation of an hyper-parameter scan for a few selected parameters.
        On the $y$-axis is represented the different values of the FID. The violin plots represent
        the behaviour of the training for a given hyper-parameter. Violins with denser tails are
    considered better choices.}
    \label{fig:hyperopt}
\end{figure*}

\begin{figure}[!h]
    \centering
    \includegraphics[width=1.01\linewidth]{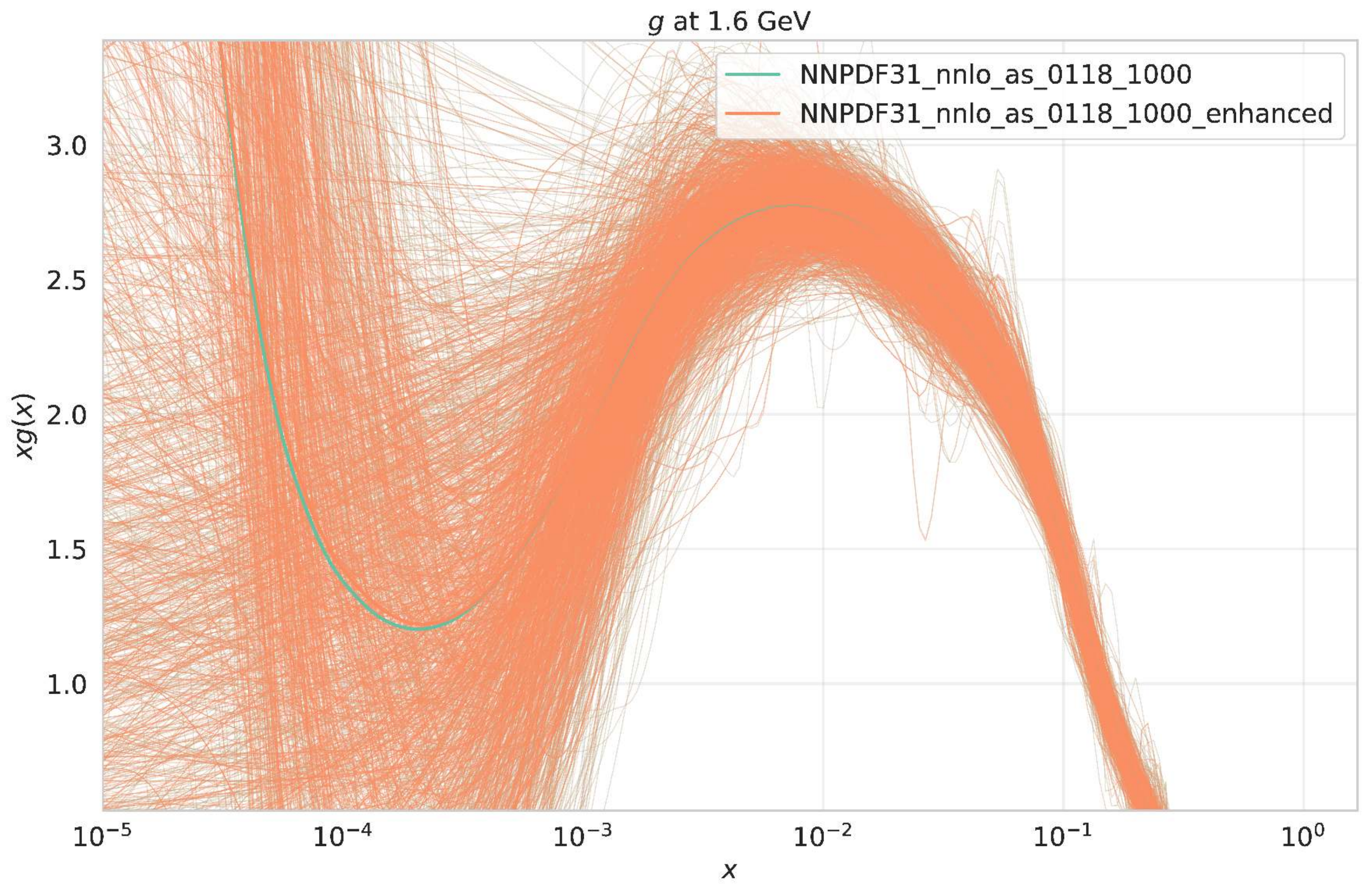}
    \caption{Comparison of the prior Monte Carlo PDF replicas with 1000 replicas and the
        synthetics with 2000 replicas generated from the prior. The synthetic sample does not
    (explicitly) contain replicas from the prior.}
    \label{fig:gluon}
\end{figure}

Throughout this paper, we always consider a PDF set with a prior of $N_p=1000$ replicas
generated using the NNPDF3.1 methodology~\cite{Ball:2017nwa}.
Plots and figures are generated using the \texttt{ReportEngine}-based \texttt{validphys}
suite~\cite{zahari_kassabov_2019_2571601}.
We use the \texttt{ganpdfs} to
generate $N_s=2000$ synthetic replicas for a total of $N_e=3000$ replicas.
In order to reduce biases, the parameters of the GAN
architecture are configured according to the results of a hyper-parameter scan.
In~\Fig{fig:hyperopt}, we plot an example of such a scan in which we
show a few selected hyper-parameters. For each hyper-parameter, the values of the FID are
plotted as a function of different parameters. The violin shapes represent a visual illustration
of how a given parameter behave during the training. That is, violins with denser tails
are considered better choices as they yield stable training. For instance, we can see that $1000$ epochs lead to more
stable results as opposed to $1500$ or $2000$. For a complete summary, the list of hyper-parameters
with the corresponding best values are shown in Table~\ref{table:architecture}.
\begin{table}[!h]
    \centering
    \begin{tabular}{c|cc}
        \hline
    & Generator & Discriminator \\
    \hline
        Network depth & 3 & 2 \\
        \hline
        kernel initializer & glorot uniform & glorot uniform \\
        \hline
        weight clipping & - & 0.01 \\
        \hline
        optimizer & None & RMSprop \\
        \hline
        activation & leakyrelu & leakyrelu \\
        \hline
        steps & 1 & 4 \\
        \hline
        batch size & \multicolumn{2}{c}{70$\%$} \\
        \hline
        number of epochs &  \multicolumn{2}{c}{1000} \\
        \hline
    \end{tabular}
    \caption{Parameters on which the hyper-parameter scan was performed are shown in the first
        column. The resulting best values are shown for both the generator and discriminator in the
    second and third column.}
    \label{table:architecture}
\end{table}
It is important to emphasize that the training of the generator has to be performed w.r.t.
the discriminator's predictions. This is the reason why no optimizer is required when training
the generator. For illustration purposes, we show in~\Fig{fig:gluon} the output of the GANs by
comparing the synthetics with the input Monte Carlo replicas.

\begin{figure}[!h]
    \centering
    \includegraphics[width=\linewidth]{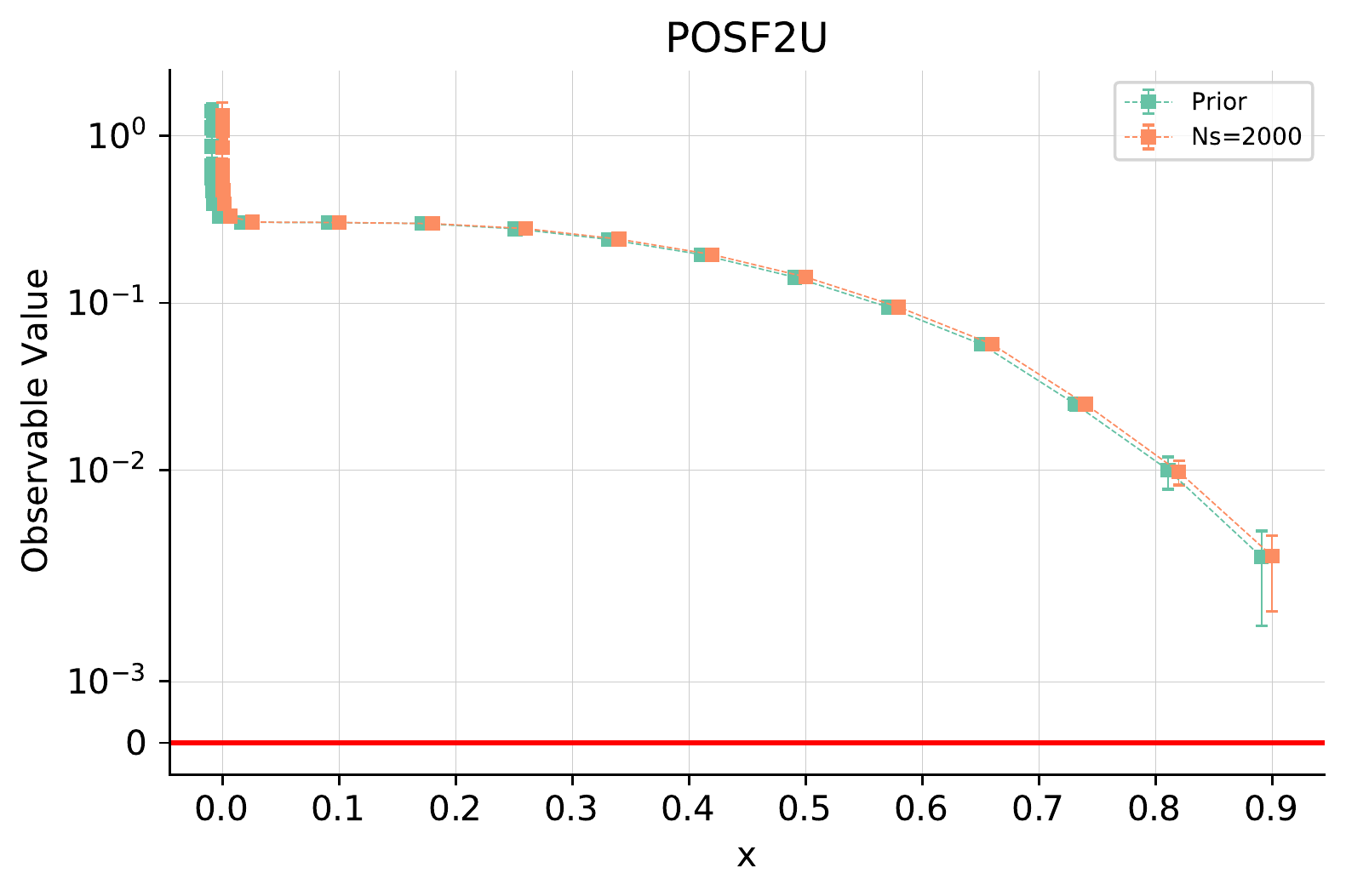}
    \caption{Positivity constraint for the prior (green) and synthetic (orange) PDF sets.
        These constraints corresponds to the positivity observable $F^{u}_{2}(x)$ introduced in Eq.~(14)
        of~\cite{Ball:2014uwa}.
    }
    \label{fig:positivity}
\end{figure}

We also verify that despite the fact that physical constraints such as sum rules and positivity
of the PDFs are not enforced when constructing the synthetic set, they are
automatically satisfied. In Table~\ref{table:sumrules} we compare the values of the sum rules
between prior and synthetic replicas. We notice that the resulting values from the synthetic
PDFs are very close to the ones from a real fit. Similar conclusions can be inferred when looking
at the positivity plots in~\Fig{fig:positivity} from which we can see that not only the positivity
of the PDFs are preserved in the synthetic PDFs, but also the results are quite close to the real
fit.
\begin{table}[!h]
	\centering
	\begin{tabular}{c|cc}
	\hline
	\textbf{Prior} & mean & std \\
	\hline
	momentum & 0.9968 & $7.315 \times 10^{-4}$ \\
	\hline
	$u_v$ & 1.985 & $3.122 \times 10^{-2}$ \\
	\hline
	$d_v$ & 0.9888 & $3.764 \times 10^{-2}$ \\
	\hline
	$s_v$ & $3.249 \times 10^{-3}$ & $3.547 \times 10^{-2}$  \\
	\hline
	\hline
	\textbf{Synthetic} & mean & std  \\
	\hline
	momentum & 0.9954 & $1.907 \times 10^{-3}$ \\
	\hline
	$u_v$ & 1.992 & $3.788 \times 10^{-2}$ \\
	\hline
	$d_v$ & 0.9956 & $3.796 \times 10^{-2}$ \\
	\hline
	$s_v$ & $2.073 \times 10^{-4}$ & $4.833 \times 10^{-2}$ \\
	\hline
	\end{tabular}
	\caption{Table comparing the values of the sum rules between the prior with $N_p=1000$ and
	synthetic PDF set with $N_s=2000$ generated from the prior using \texttt{ganpdfs}.}
	\label{table:sumrules}
\end{table}

\section{Results}
\label{sec:results}

In this section, we quantify the performance of the GAN-enhanced compression methodology
described in the previous section based on various statistical estimators. First, as a validity check, we compare
the central values and luminosities of the GAN-enhanced compressed sets with the results from
the original PDF set and the standard compression. Then, in order to estimate how good the
GAN-compressor framework is compared to the previous methodology, we subject the compressed
sets resulting from both methodologies to more visual statistical estimators such
as correlations between different PDF flavours. Here, we consider the same Monte Carlo PDF
sets as the ones mentioned in the previous section, namely a prior set with $N_p=1000$
replicas which was enhanced using \texttt{ganpdfs} to
$N_e=3000$ replicas (i.e., $N_s=2000$ synthetic replicas). In all the cases, the compression of
the PDF sets are handled by the \texttt{pyCompressor} code.

\subsection{Validation of the GAN-compressor}

\begin{figure}[!h]
	\centering
	\includegraphics[width=\linewidth]{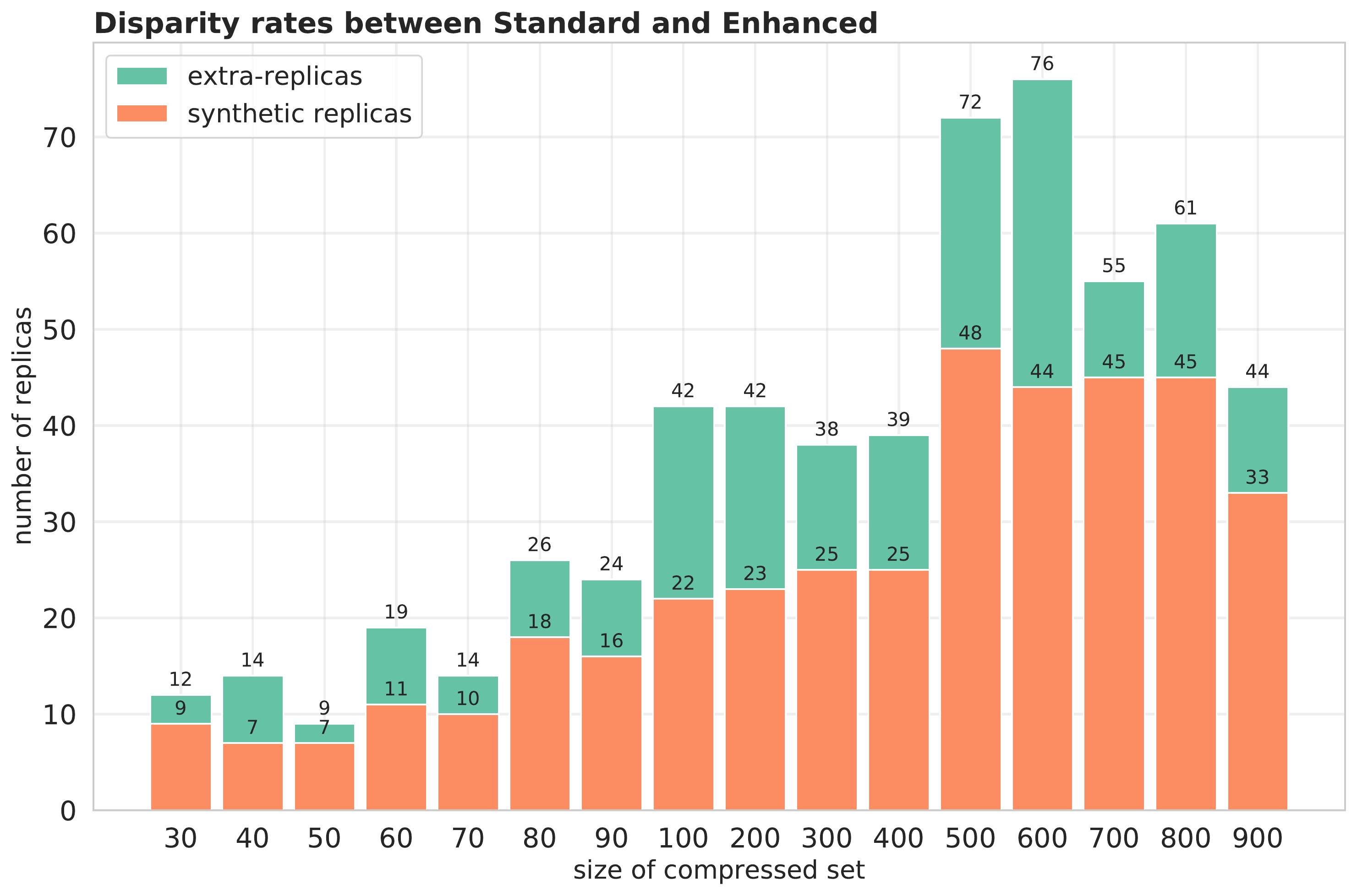}
	\caption{The histograms represent the number of disjoint replicas that are present in the
	standard compressed set but not in the GAN-enhanced for the NNPDF4.0 candidate. The results
	are shown as a function of the size of the compressed set. The numbers of synthetic replicas
	are included.}
	\label{fig:disparity}
\end{figure}

First, we would like to see, for a given compression from the enhanced set, how many replicas
are selected from the synthetic set. We consider the compression of the enhanced set
into subsets with smaller number of replicas. In~\Fig{fig:disparity}, we show the disparity
rate between the standard and GAN-enhanced compressed sets, i.e. the number of replicas that
are present in the GAN-enhanced compressed sets (including synthetics) but not in the standard
sets. We also highlight the number of synthetic replicas that end up in the final set.
The results are shown for various sizes of the compressed set. For smaller sizes
(smaller than $N_c=200$), the percentage of synthetic replicas exceeds
$10\%$ and this percentage decreases as $N_c$ increases. This is explained by the fact
that as the size of the compressed set approaches the size of the input PDF, the probability
distribution of the reduced real samples get closer to the prior and fewer synthetics are
required.

\begin{figure*}[!ht]
	\captionsetup[subfigure]{aboveskip=-1.5pt,belowskip=-1.5pt}
	\centering
	\begin{subfigure}{0.385\linewidth}
		\includegraphics[width=\linewidth]{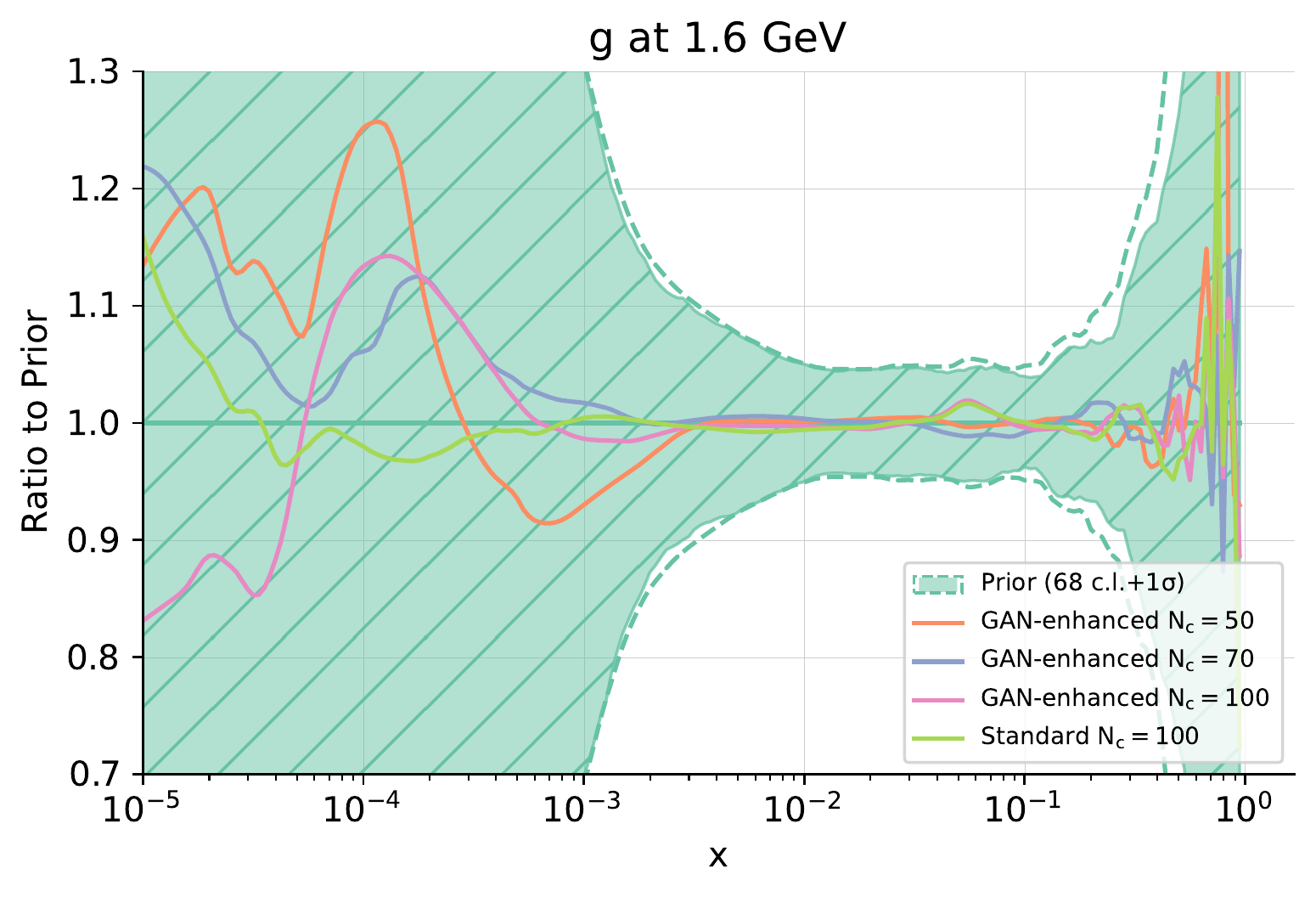}
	\end{subfigure}
	\hfil
	\begin{subfigure}{0.385\linewidth}
		\includegraphics[width=\linewidth]{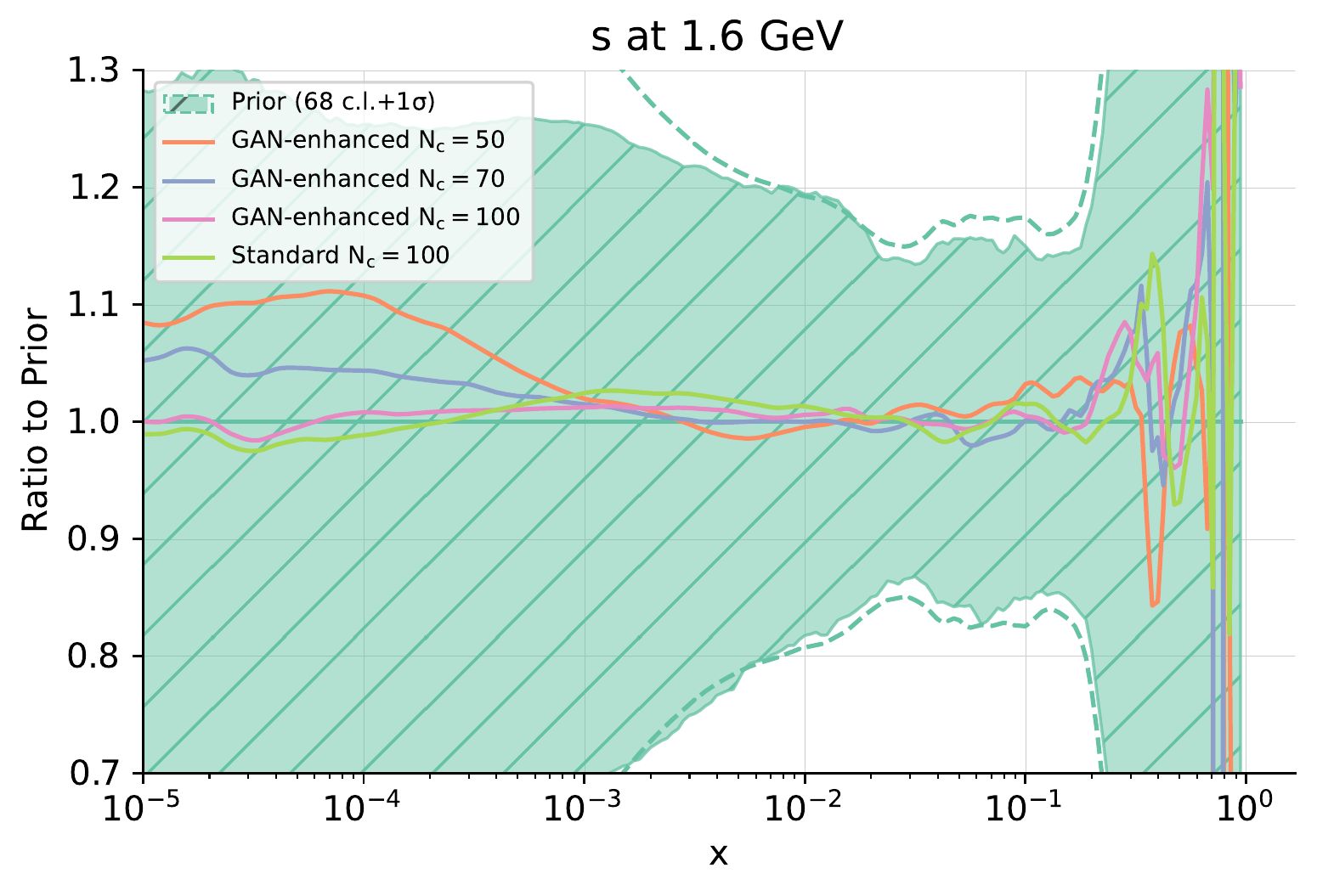}
	\end{subfigure}
	\begin{subfigure}{0.385\linewidth}
		\includegraphics[width=\linewidth]{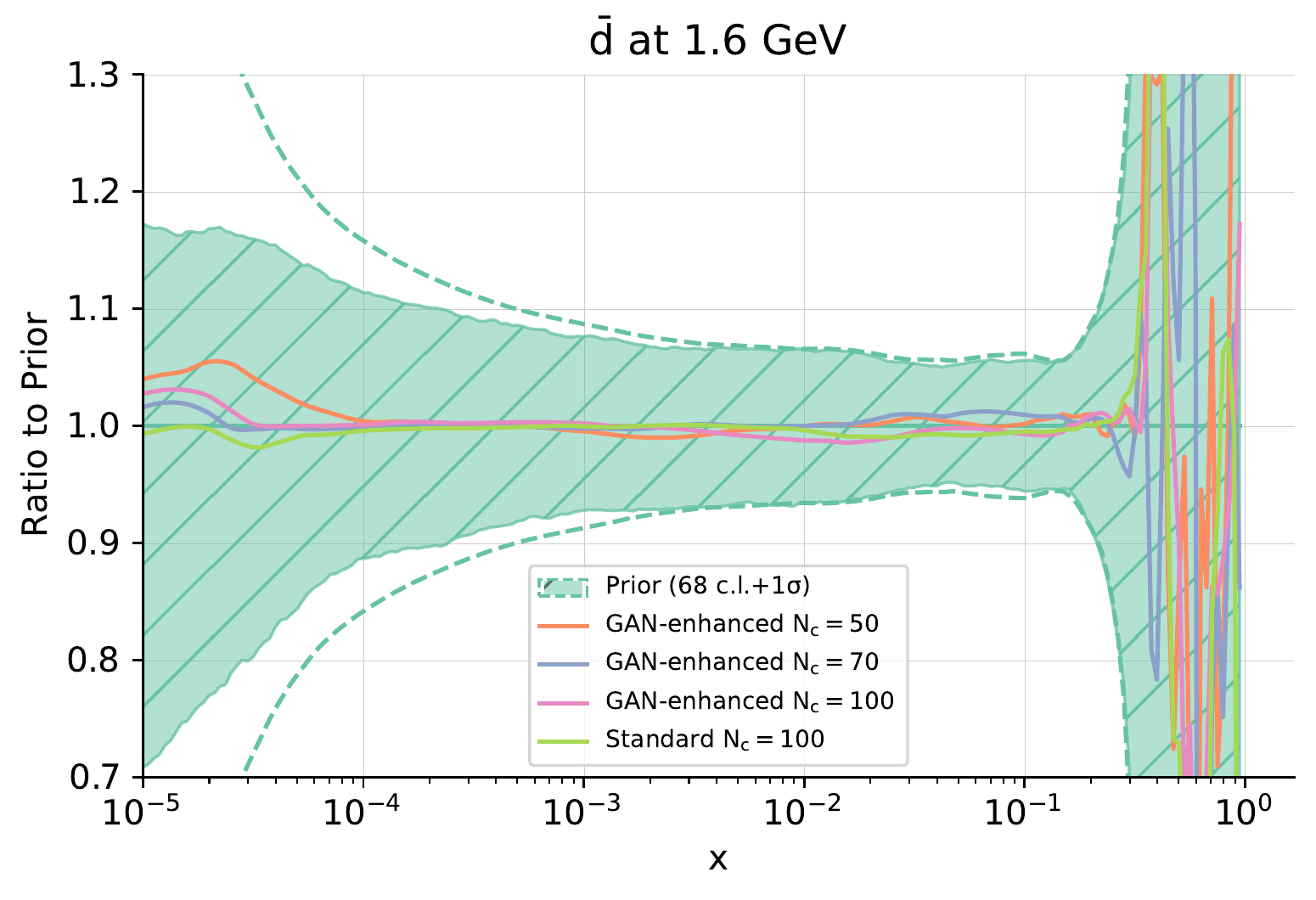}
	\end{subfigure}
	\hfil
	\begin{subfigure}{0.385\linewidth}
		\includegraphics[width=\linewidth]{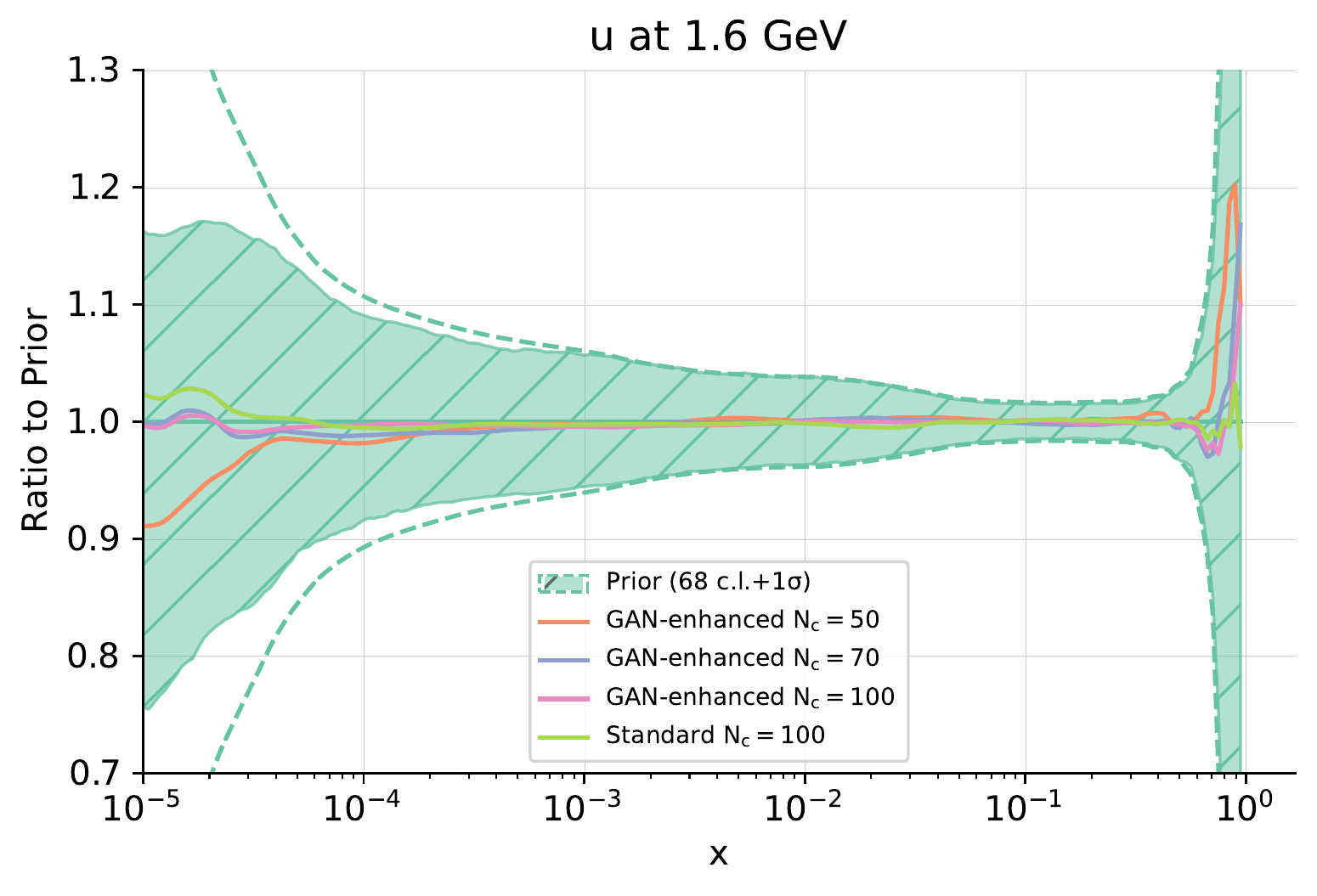}
	\end{subfigure}
	\caption{Comparison of the PDF central values resulting from the GAN-enhanced compression for
		different values of $N_c$ $=50, 70, 100 $ and different PDF flavours $(g, s, \bar{d}, \bar{s})$.
		Results are normalized to the central value of the prior set. The $68\%$ c.l. is represented by the
		green hatched band while the 1-sigma is given by the dashed green curves. As a reference, results from
		the standard compression for $N_c=100$ are also shown.}
	\label{fig:cv}
\end{figure*}

\begin{figure*}[!h]
	\captionsetup[subfigure]{aboveskip=-1.5pt,belowskip=-1.5pt}
	\centering
	\begin{subfigure}{0.385\linewidth}
		\includegraphics[width=\linewidth]{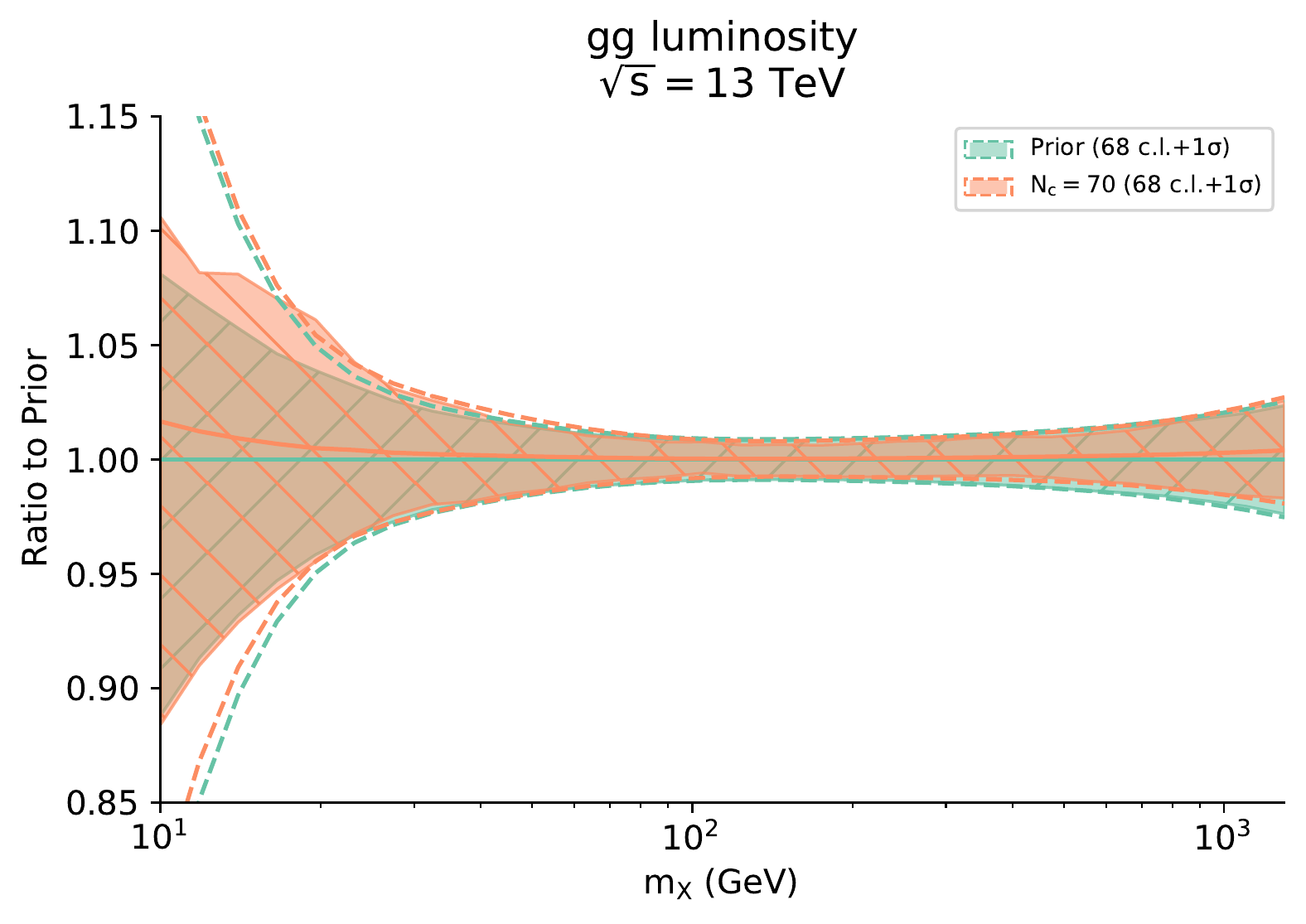}
	\end{subfigure}
	\hfil
	\begin{subfigure}{0.385\linewidth}
		\includegraphics[width=\linewidth]{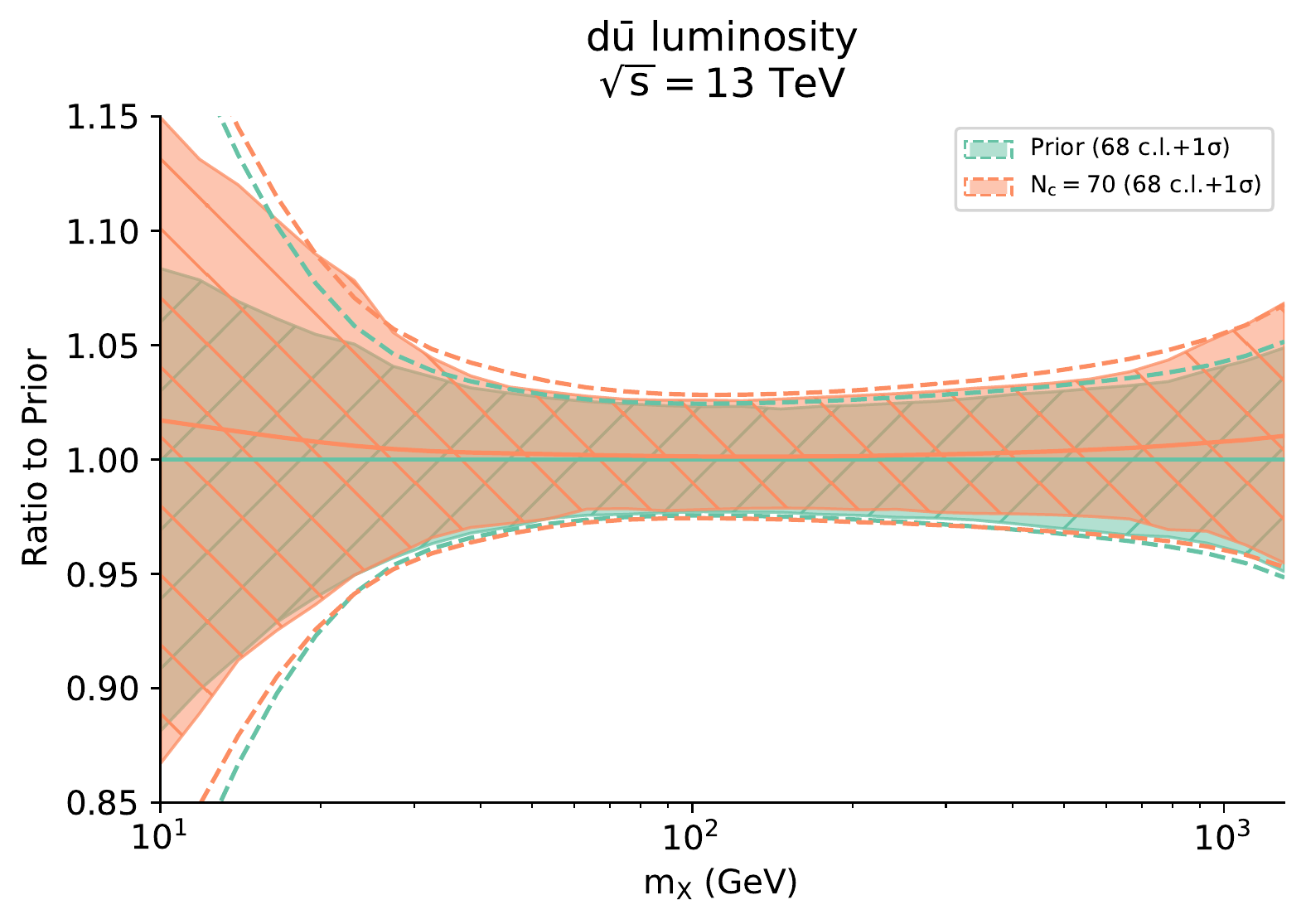}
	\end{subfigure}
	\begin{subfigure}{0.385\linewidth}
		\includegraphics[width=\linewidth]{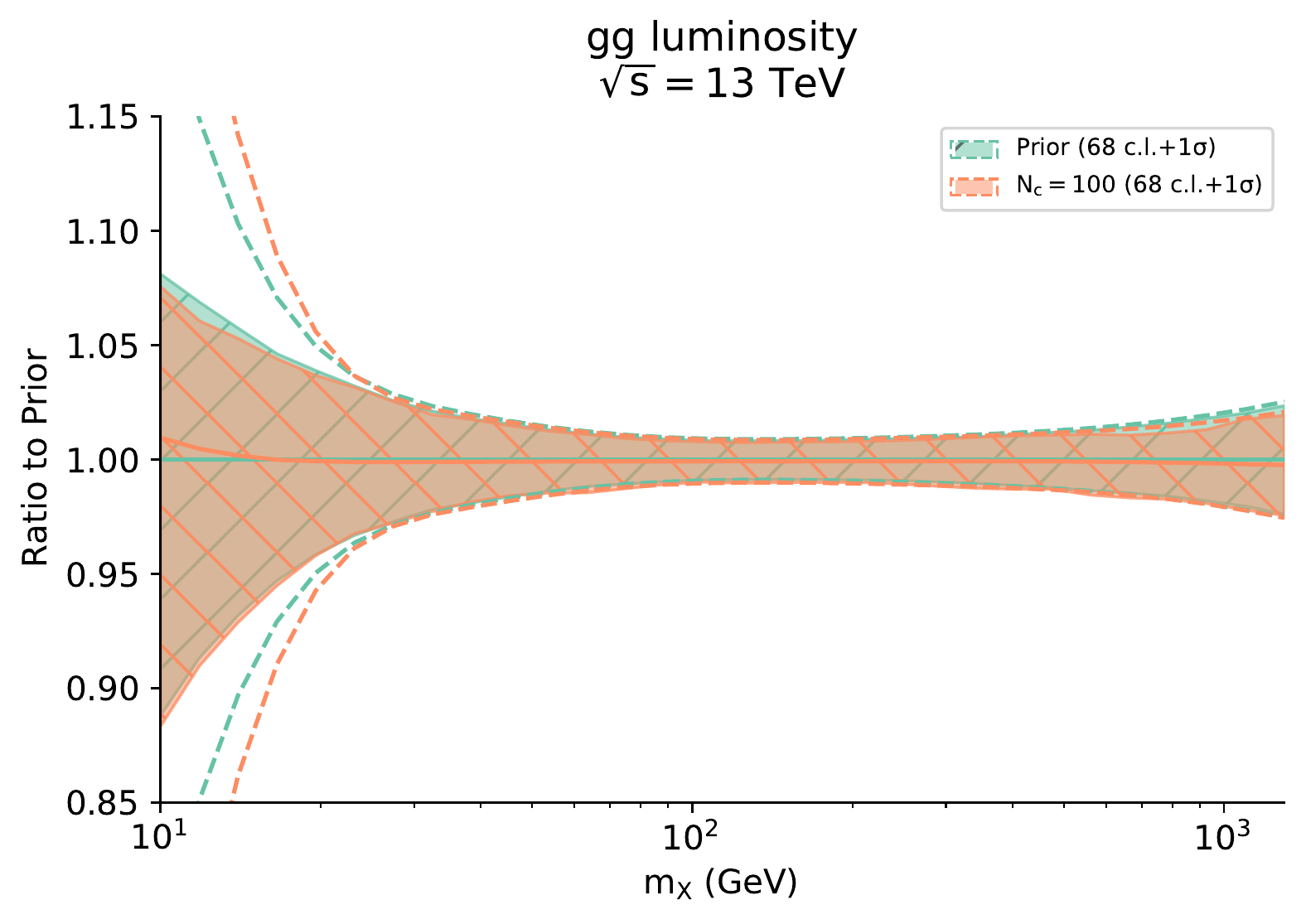}
	\end{subfigure}
	\hfil
	\begin{subfigure}{0.385\linewidth}
		\includegraphics[width=\linewidth]{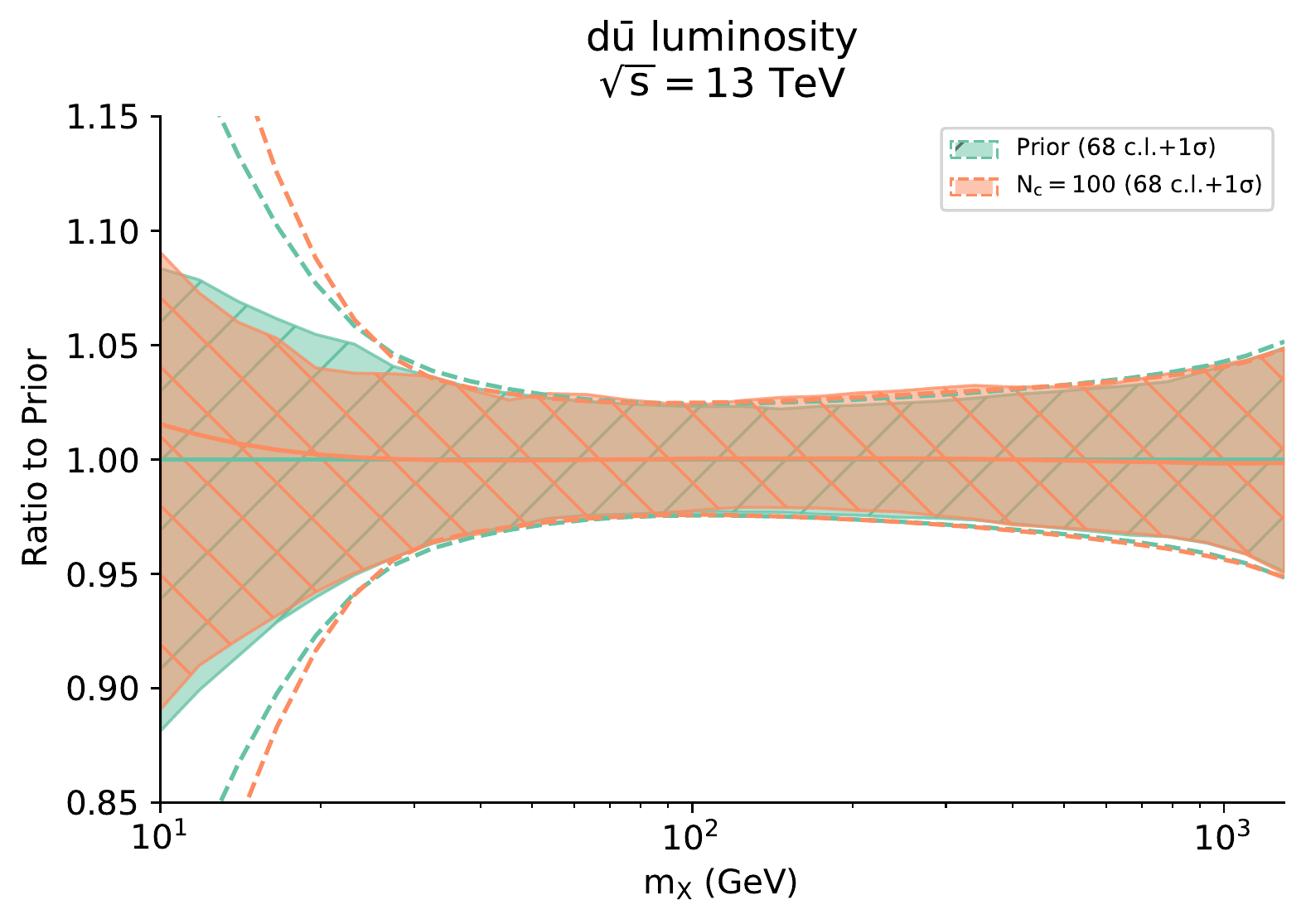}
	\end{subfigure}
	\caption{Comparison of the PDF luminosities between the prior and the GAN-enhanced compressed
		sets at LHC with $\sqrt{s} = \gev{13}$. The results are shown for different $N_c = 70, 100$ and
		for different PDF luminosities ($gg$ and $d \bar{u}$). The error bands represent the $68\%$ c.l.}
	\label{fig:luminosity}
\end{figure*}

Now we turn to the validation of the GAN-enhanced compression methodology by looking
at the PDF central values and luminosities. In~\Fig{fig:cv}, we plot the absolute central
values of the reduced sets resulting from the compression of the GAN-enhanced Monte Carlo
replicas at a scale $Q = \gev{1.65}$. The results are presented for three different sizes of
compressed sets $N_c = 50, 70, 100$ and for various PDF flavours $(g, s, \bar{d}, \bar{s})$.
The results are normalized to the central value of the prior Monte Carlo PDF replicas.
In order to quantify whether or not samples of $N_c= 50, 70, 100$ are good representations
of the probability distributions of the prior PDF replicas, we plot the $68\%$ c.l. and the
1-sigma band.
We see that the PDF uncertainties are much larger than the fluctuations of the central values,
indicating that a compressed set with size $N_c = 50$ captures the main statistical
properties of the prior.
In~\Fig{fig:luminosity} we plot the luminosities for the $g$-$g$, $d$-$\bar{u}$ combinations
as a function of the invariant mass of the parton pair $M_{x}$ for two different
compressed sizes: $N_c = 70, 100$.
The error bands represent the 1-sigma confidence interval.
At lower values of $M_X$ where PDFs are known to be non-Gaussian, the $N_c =70$ compressed set slightly
deviate from the underlying probability distributions. However, the deviations are very
small compared to the uncertainty bands. For $N_c = 100$, we see very good agreement between
the prior and the compressed set.

The above results confirm that compressed sets extracted from the GAN-enhanced compression
framework fully preserve the PDF central values and luminosities. In particular, we conclude
that about $50$ replicas are sufficiently enough to reproduce the main statistical properties
of an input with $N_p = 1000$ replicas. Next, we quantity how efficient the generative-based
compressor strategy is compared to the standard approach.

\subsection{Performance of the GAN-enhanced compressor}

In order to quantify how good the GAN-compressor framework is compared to the previous
methodology, we evaluate the compressed sets resulting from both methodologies on various
statistical indicators. We consider the same settings as in the previous sections, namely
a prior with $N_p=1000$ enhanced with the GAN to generate $N_s=2000$ synthetic replicas.
The results from the GAN-enhanced compressor are then compared to the results from the
standard compression in which the subset of replicas are selected directly from the prior.
\begin{figure*}[!h]
    \centering
    \includegraphics[width=1.0\textwidth]{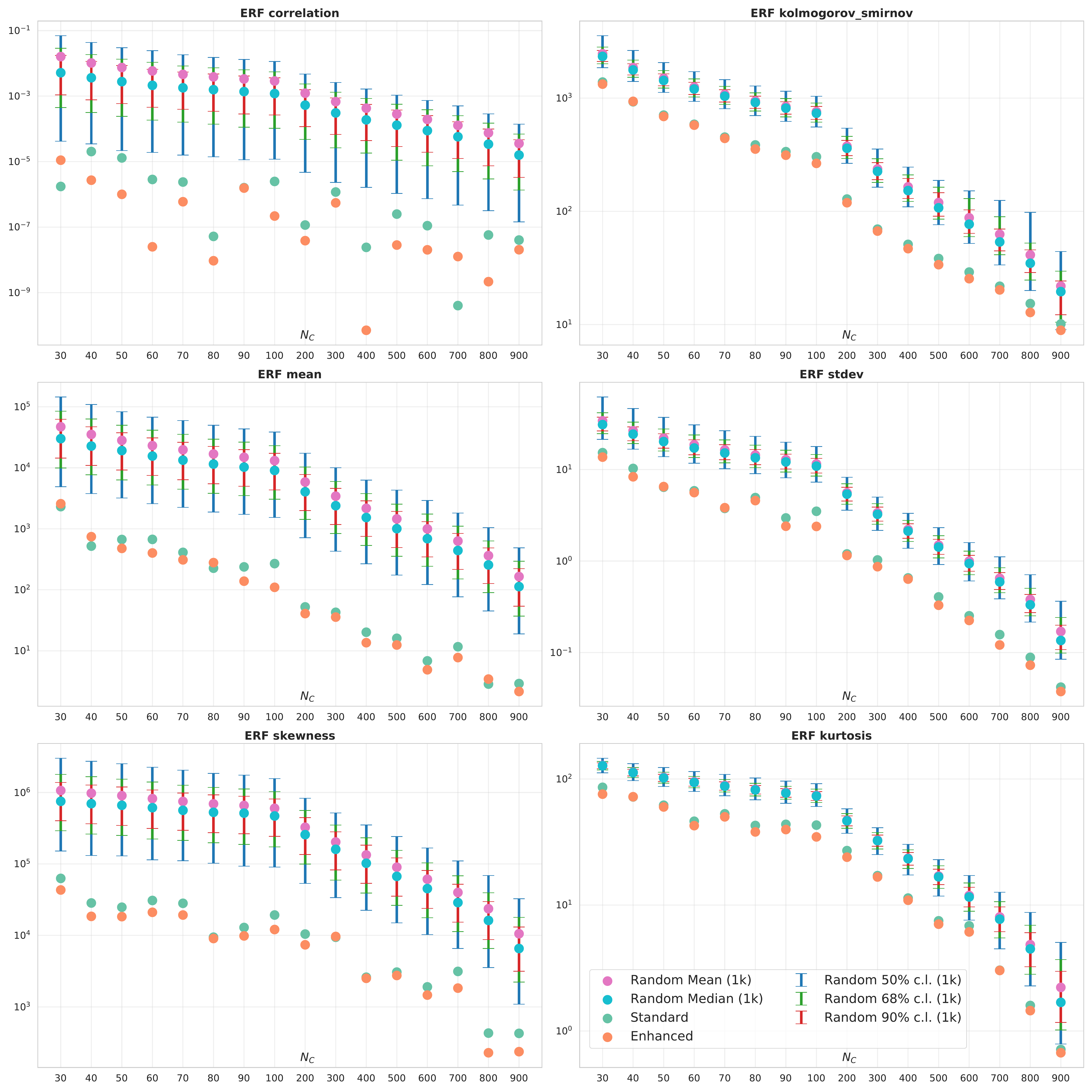}
    \caption{Comparison of the best ERF values for the compression of a Monte Carlo PDF with
        $N_p=1000$ replicas. For each compressed set, we show the contribution of each of the statistical
        estimators that contribute to the total ERF using the standard compression (green) and the
        GAN-enhanced compression (orange) methodology. For illustration purposes, the mean (purple)
        and median (light blue) resulting from the average of $N_r =1000$ random selections are shown.
        The resulting confidence intervals from the random selections are represented by the blue $(50\%)$,
    green $(68 \%)$, and red $(90 \%)$ error bars.}
    \label{fig:erfs}
\end{figure*}

In~\Fig{fig:erfs}, for each
compressed set, we show the contribution of each statistical estimators (mean, standard
deviation, Kurtosis, Skewness, Kolmogorov distance, and correlation) that contribute to the
total value of the ERF using the standard (green) and GAN-based (orange) approach
as a function of the size of the compressed set.
For reference, we also show the mean (purple) and median (light blue) computed by taking the
average ERF values from $N_r = 1000$ random selections. The confidence intervals ($50\%, 68\%, 90\%$)
computed from the random selections are shown as error bars of varying colours and provide an
estimate of how representative of the prior distribution a given compressed set is.

First of all, we see from the plots that as the size of the compressed set increases,
the ERF values for all the estimators tend to zero. On the other hand, it is clear
that both compression methodologies outperform quite significantly any random selection of
replicas.
But in addition, by comparing the results from the standard and GAN-enhanced
compression we observe that the estimators for the enhanced compression are, in all cases except
for a very few, below those of the standard compression.
This suggests that the GAN-enhanced approach will result in a total value of ERF that is much
smaller than the one from the standard compression methodology.

\begin{figure}[!ht]
    \centering
    \includegraphics[width=\linewidth]{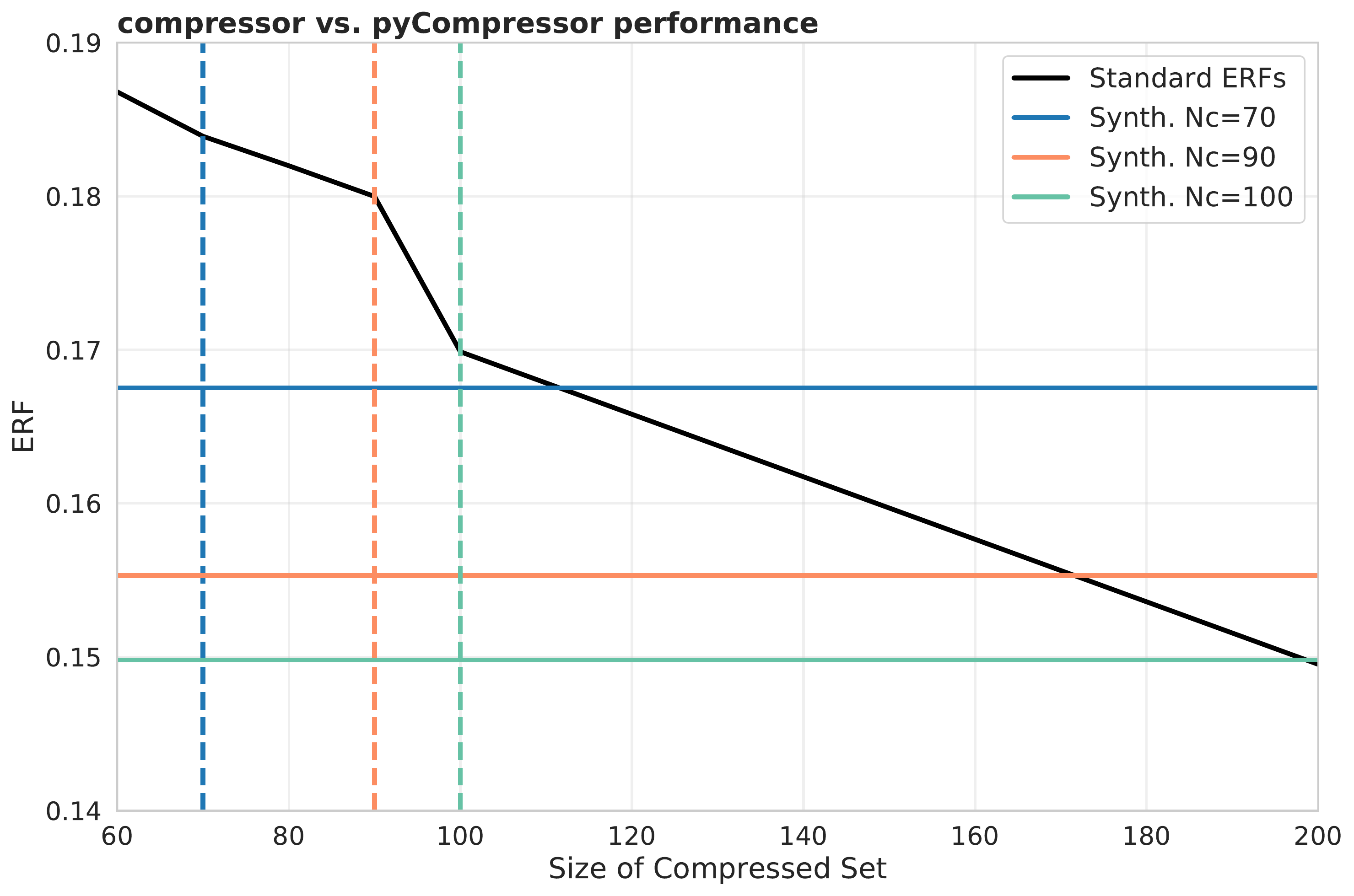}
    \caption{Comparison of the performance of the new generative-based compression algorithm (synthetic)
        and the the previous methodology (standard). The normalized ERF values of the standard compression
        are plotted as a function of the size $N_c$ of the compressed set (solid black line). The dashed blue,
        orange, and green lines represent $N_c =70, 90, 100$ respectively. The solid lines represent the
    corresponding ERF values of the enhanced compression.}
    \label{fig:performance}
\end{figure}

In terms of efficiency, these results imply that the GAN-enhanced methodology outperforms the
standard compression approach by providing a more adequate representation of the probability
distribution of the prior. This is illustrated in~\Fig{fig:performance}
in which we plot in solid black line the total ERF values for the standard compression as
a function of the size of the compressed set. The vertical dashed lines represent the size
of the compressed set $N_c=70, 90, 100$ while the horizontal solid lines represent the
respective ERF values for the generative-based compression. The intersection between vertical
dashed lines and horizontal solid lines below the black line indicate that the GAN-based compression
outperform the standard approach. For instance, we see that $N_c=70$ from the enhanced compressed set
is equivalent to about $N_c=110$ from the standard compression, and $N_c=90$ from the enhanced
compressed set provide slightly more statistics than $N_c=150$ from the standard compression.

\begin{figure*}[!h]
    \captionsetup[subfigure]{aboveskip=-1.5pt,belowskip=-1.5pt}
    \centering
    \begin{subfigure}{0.375\linewidth}
        \includegraphics[width=\linewidth]{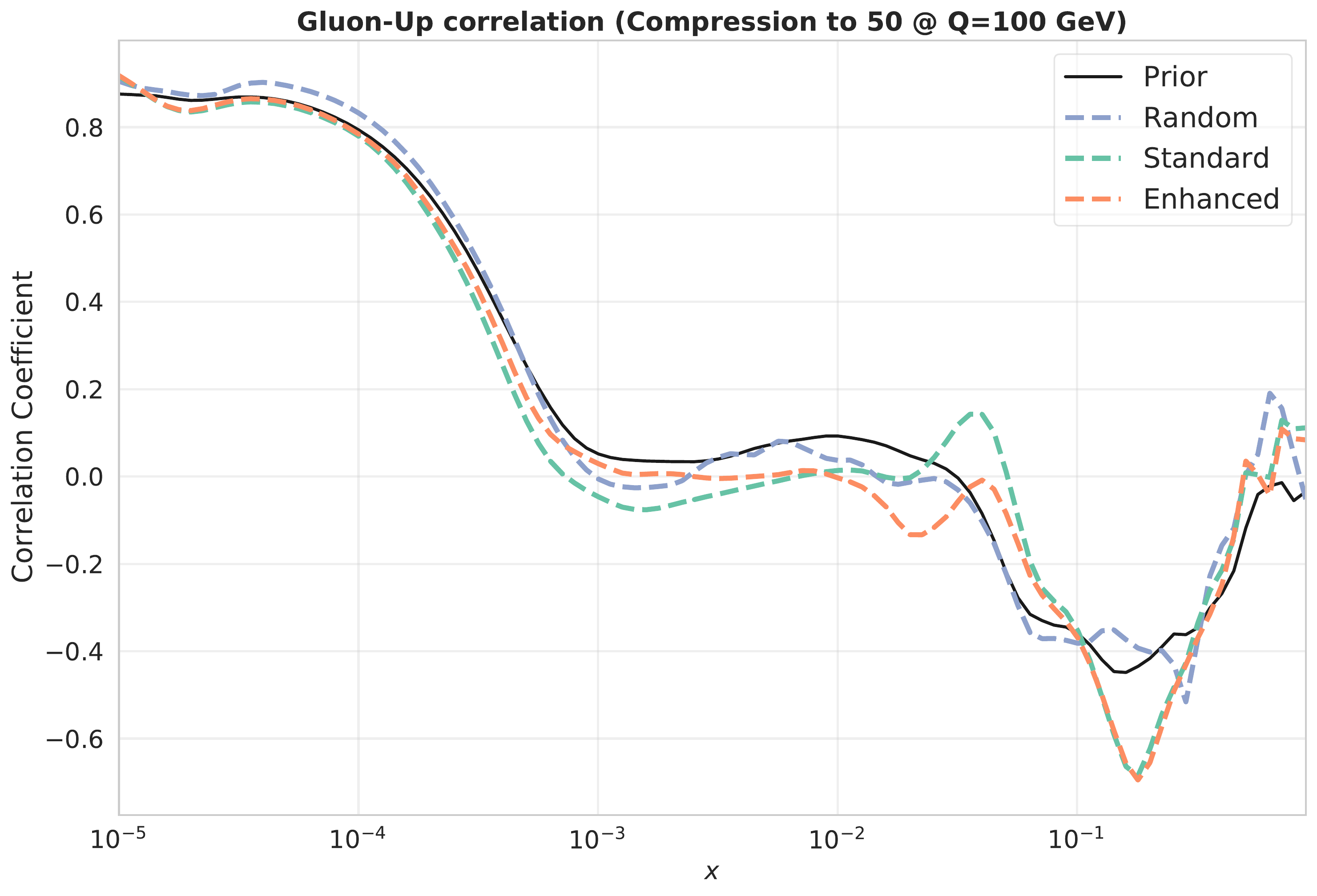}
    \end{subfigure}
    \hfil
    \begin{subfigure}{0.375\linewidth}
        \includegraphics[width=\linewidth]{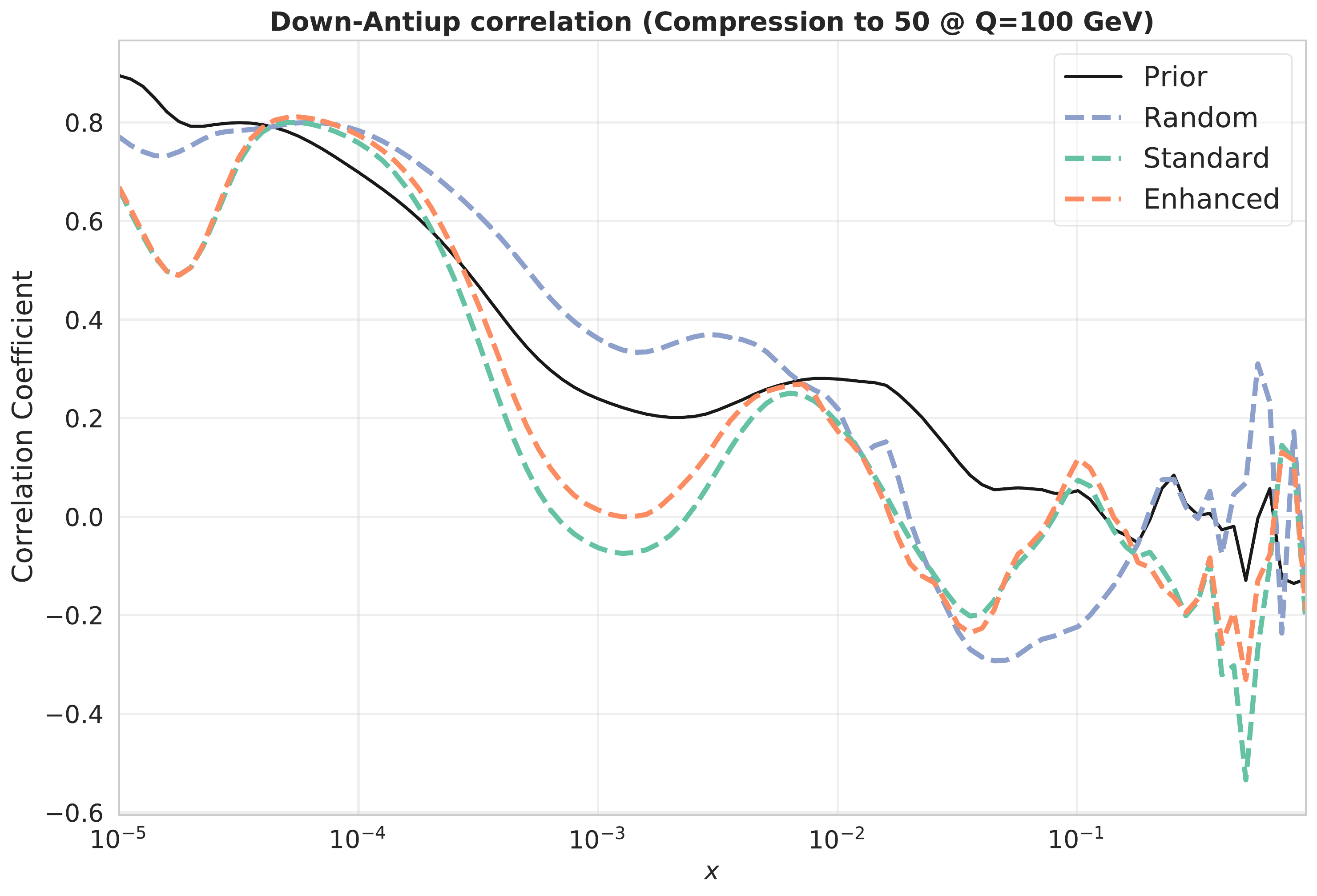}
    \end{subfigure}
    \begin{subfigure}{0.375\linewidth}
        \includegraphics[width=\linewidth]{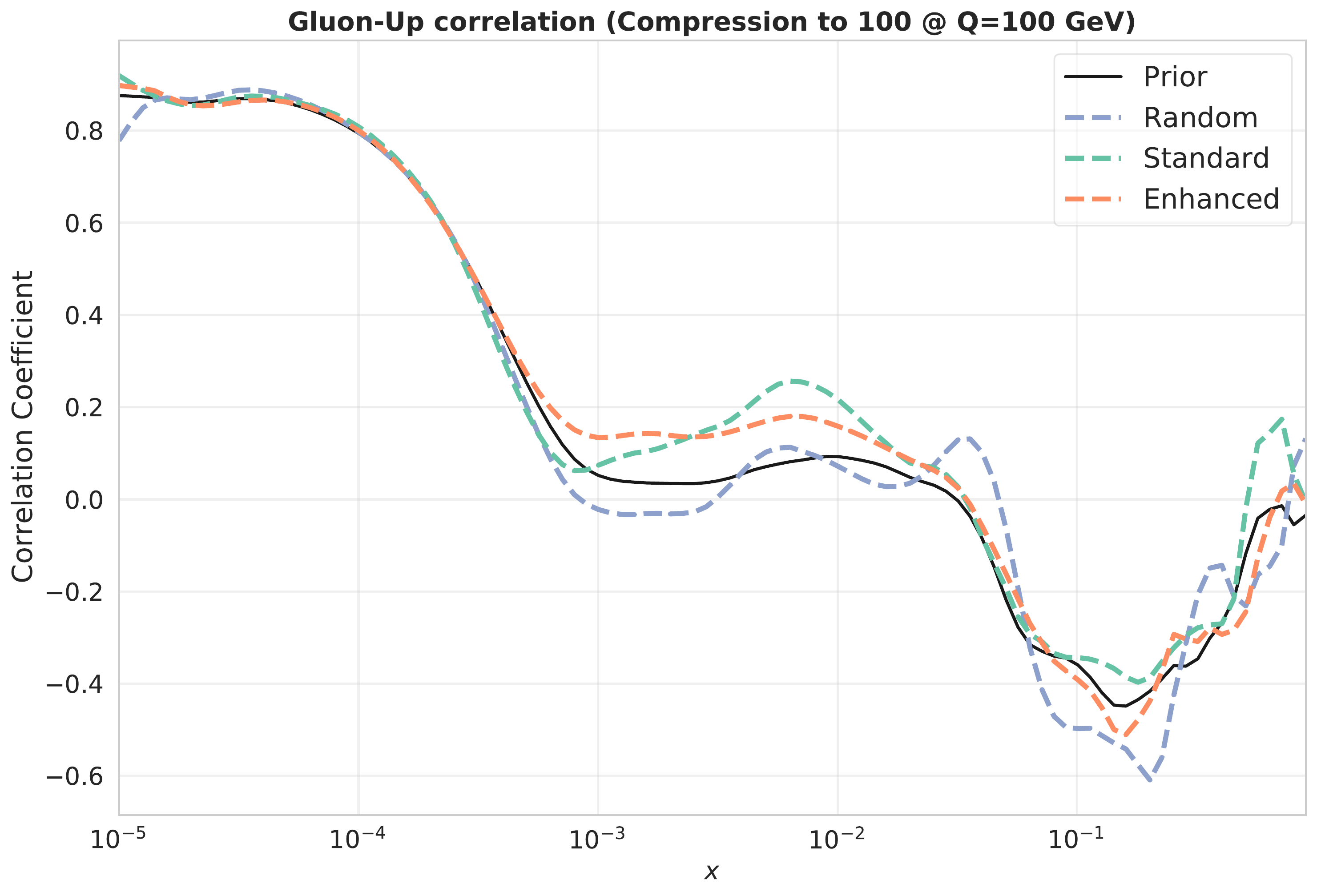}
    \end{subfigure}
    \hfil
    \begin{subfigure}{0.375\linewidth}
        \includegraphics[width=\linewidth]{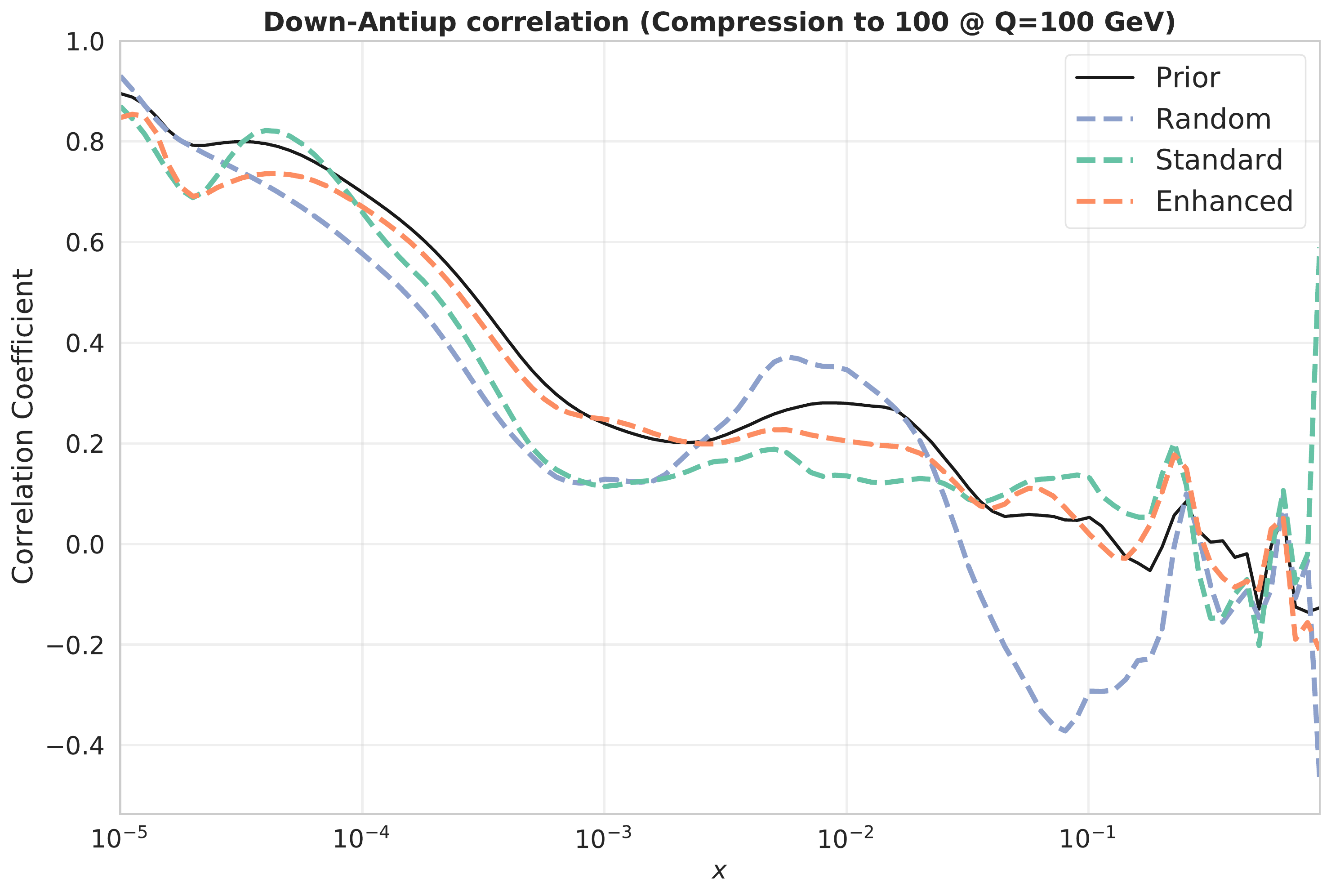}
    \end{subfigure}
    \caption{Correlations between $g$-$u$ and $d$-$\bar{u}$ PDFs for compressed sets with size
        $N_c=50$ (top) and $N_c=100$ (bottom). The energy scale $Q$ has been chosen to be $Q=\gev{100}$.
        The correlation extracted from the results of the GAN-compressor (orange) is compared to the
    results from the standard compressor (green).}
    \label{fig:correlations}
\end{figure*}

In addition to the above checks, one can also verify that correlations between PDFs are well
preserved after the compression. It is important to emphasize that one of the main differences
between a fit with $100$ and $1000$ Monte Carlo replicas is that correlations are reproduced
more accurately in the latter~\cite{Carrazza:2015hva}. This is one of the main reasons why the compression methodology
is important. Here, we show that for the same size of compressed set, the resulting compression
from the GAN-enhanced methodology also reproduces more accurately the correlations from the prior
than the standard compression. One way of checking this is to plot the correlations between
two given PDFs as a function of the Bjorken variable $x$. In~\Fig{fig:correlations}, we show
the correlation between a few selected pairs of PDFs ($g$-$u$ and $d$-$\bar{u}$) for $N_c=50, 100$
at an energy scale $Q=\gev{100}$. The results from the GAN-enhanced compression (orange) are
compared to the ones from the standard approach (green). For illustration purposes, we also
show PDF correlations from sets of randomly chosen replicas (dashed black lines). We see that
both compression methodologies capture very well the PDF correlations of the prior distribution.
Specifically, in the case $N_c=100$, we see small noticeable differences between the old and
new approach, with the new approach approximating best the original.
\begin{figure*}[!h]
    \captionsetup[subfigure]{aboveskip=-1.5pt,belowskip=-1.5pt}
    \centering
    \begin{subfigure}{0.3\linewidth}
        \includegraphics[width=\linewidth]{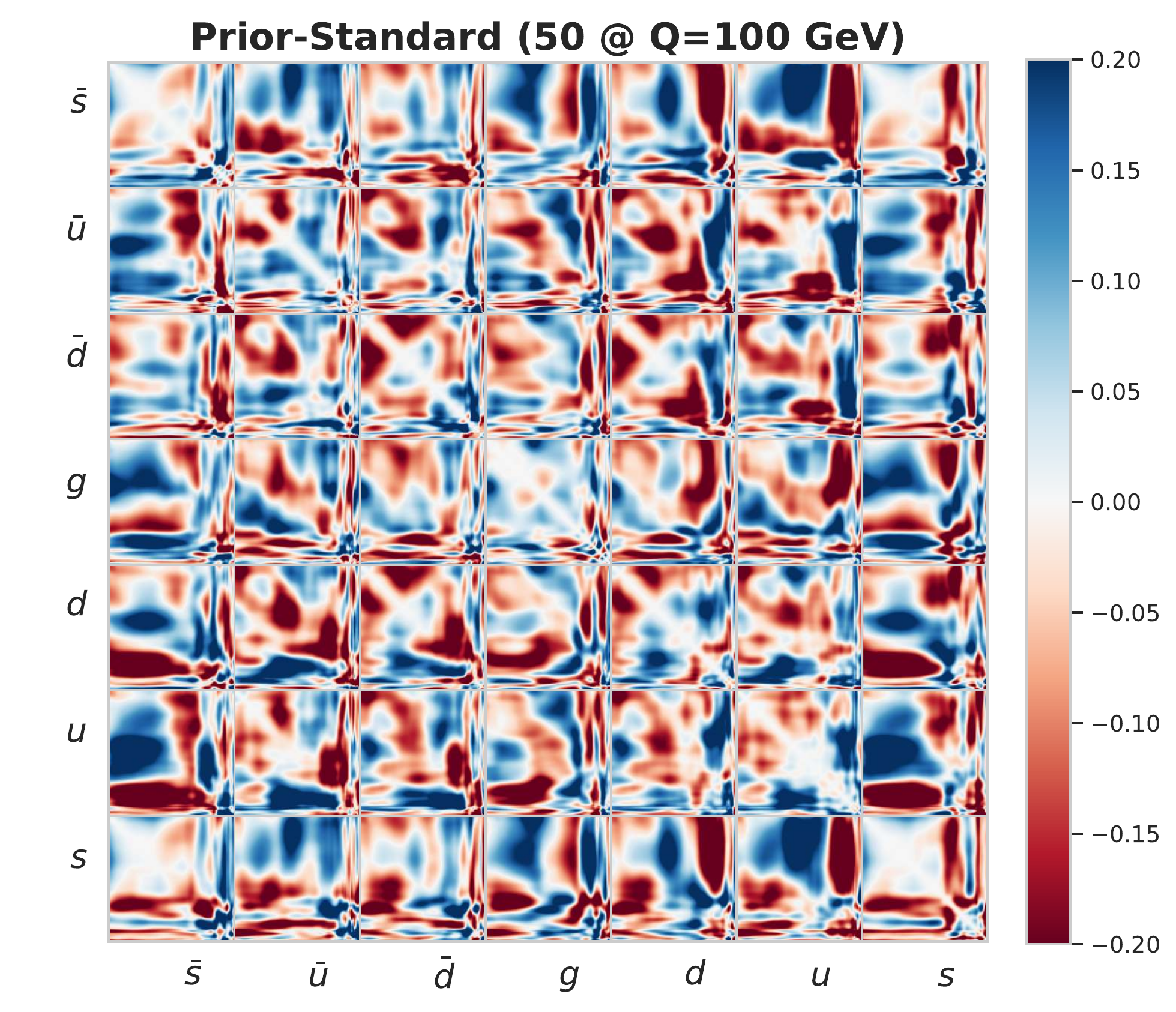}
    \end{subfigure}
    \begin{subfigure}{0.3\linewidth}
        \includegraphics[width=\linewidth]{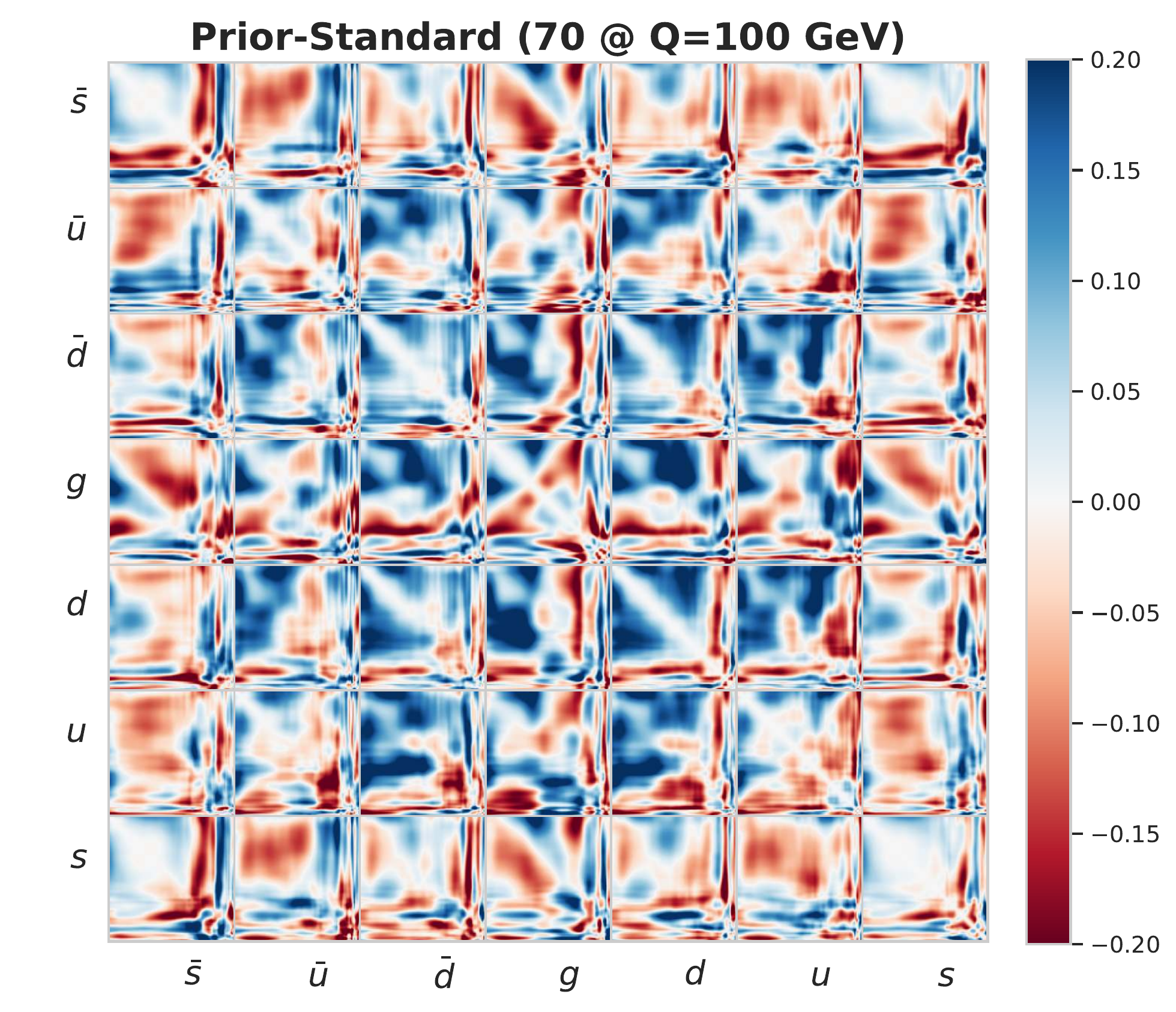}
    \end{subfigure}
    \begin{subfigure}{0.3\linewidth}
        \includegraphics[width=\linewidth]{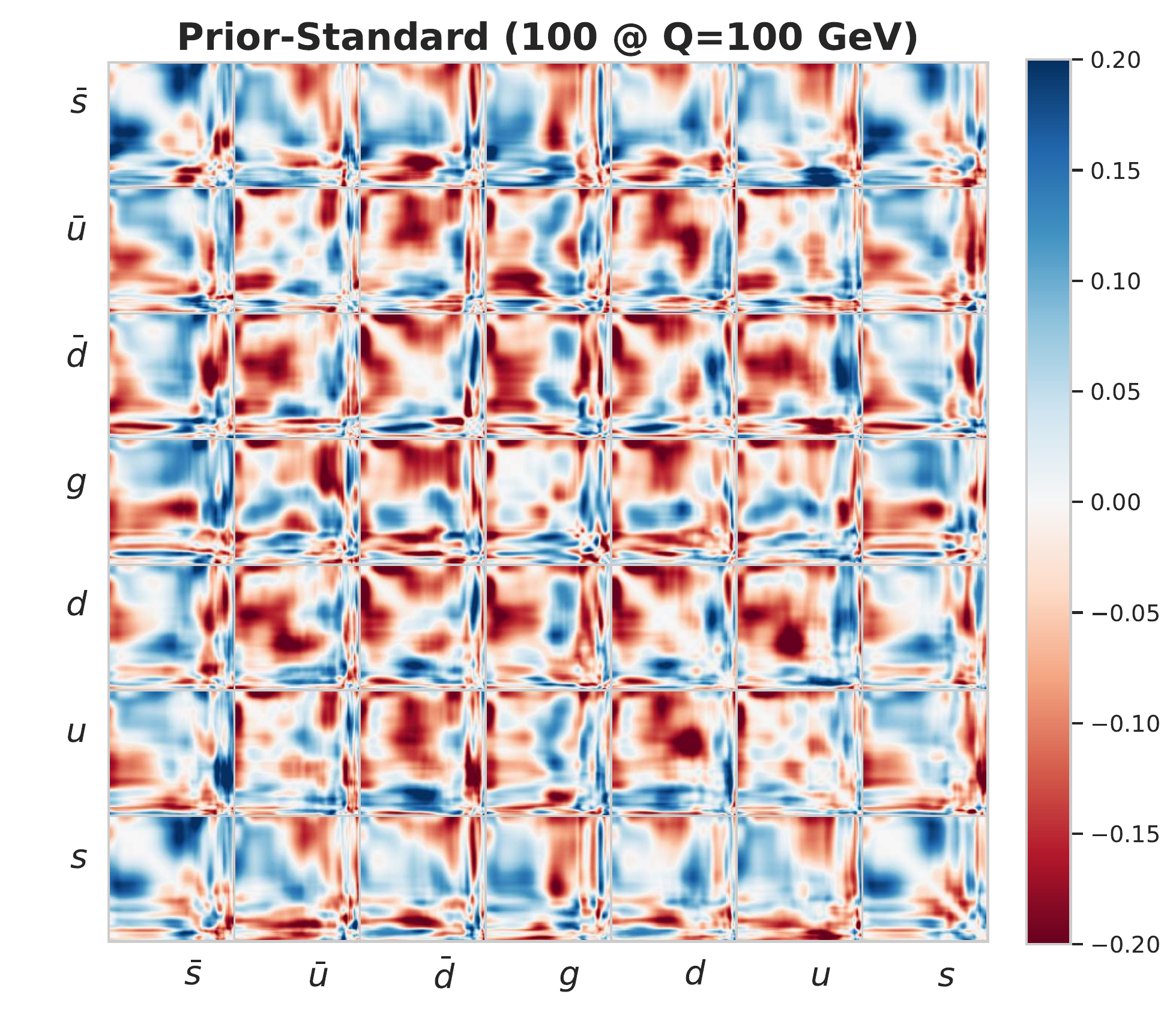}
    \end{subfigure}
    \begin{subfigure}{0.3\hsize}
        \includegraphics[width=\linewidth]{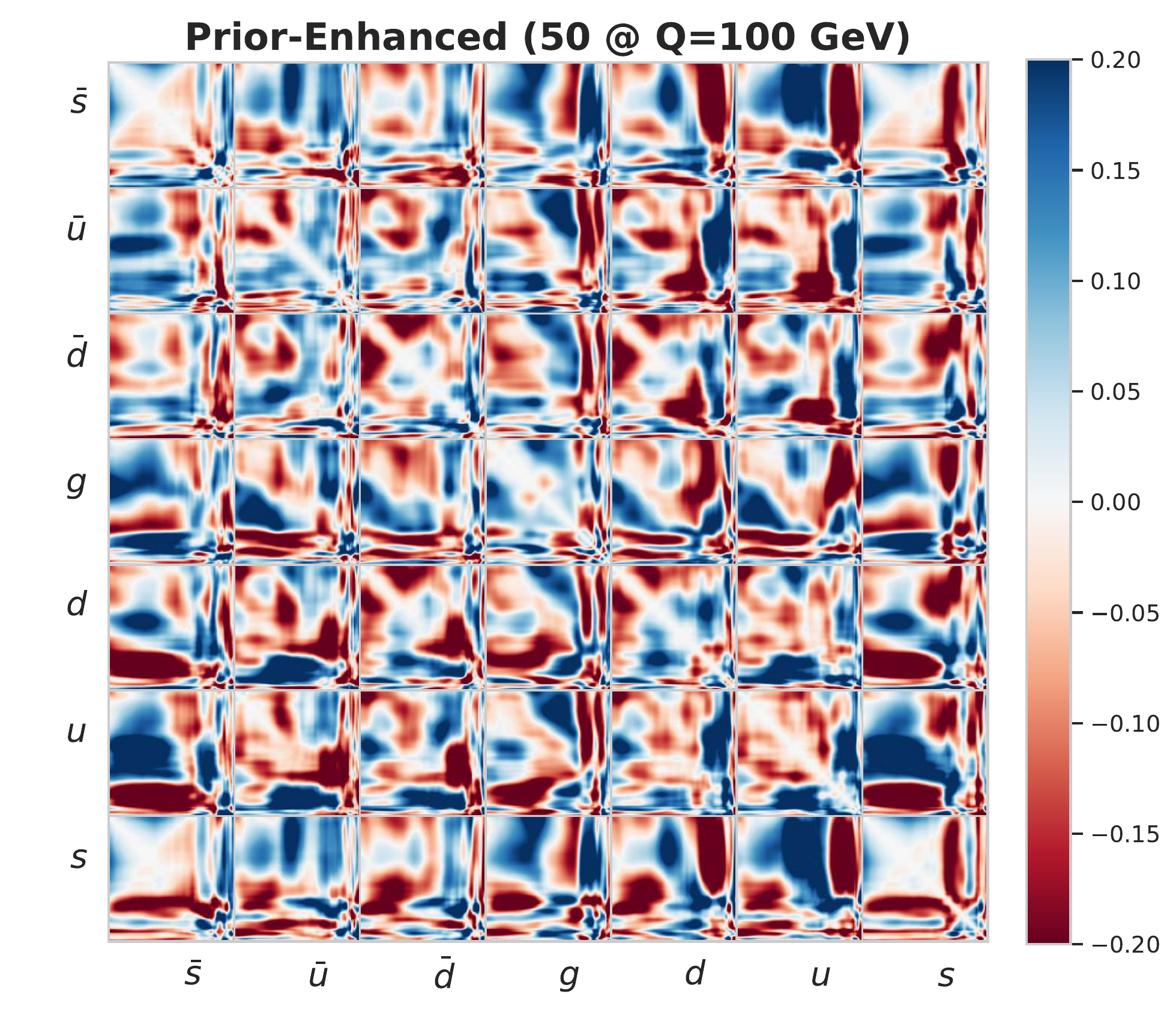}
    \end{subfigure}
    \begin{subfigure}{0.3\linewidth}
        \includegraphics[width=\linewidth]{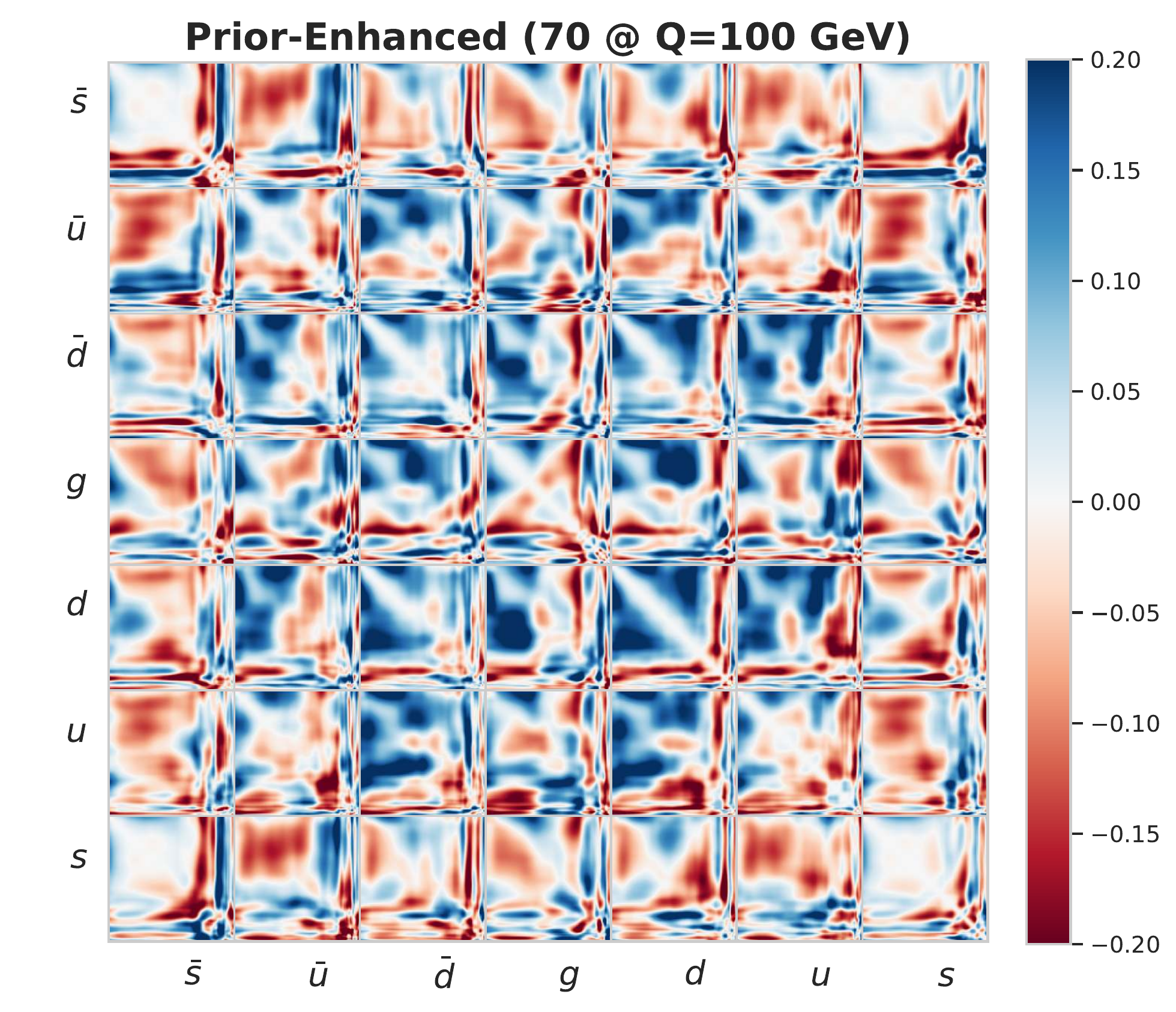}
    \end{subfigure}
    \begin{subfigure}{0.3\hsize}
        \includegraphics[width=\linewidth]{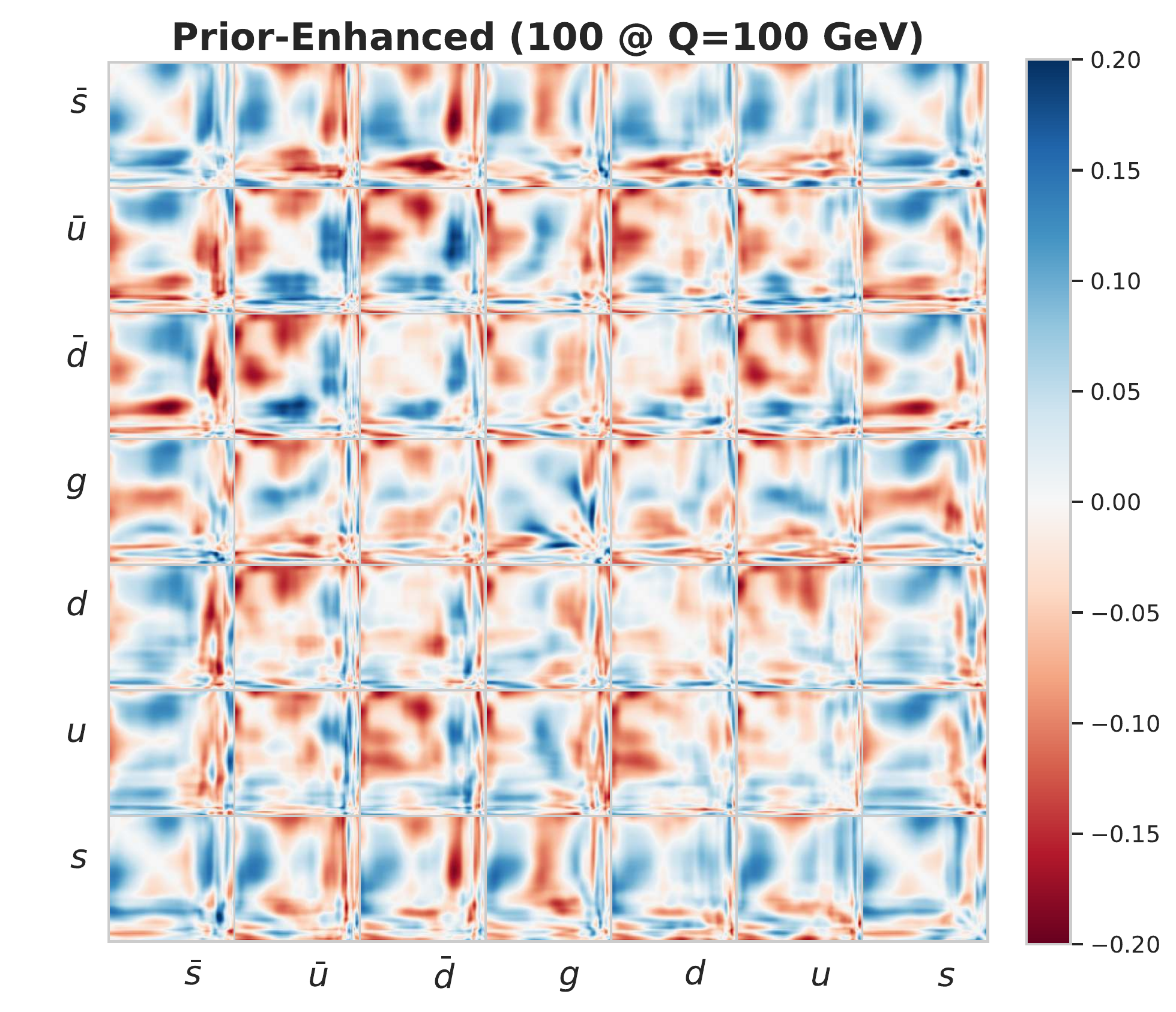}
    \end{subfigure}
    \caption{Difference between the correlation matrices of the prior and the compressed set
        resulting from the standard (first row) or enhanced (second row) compression. The correlation
        matrices are shown for different sizes of the compressed set: $N_c=50$ (first column), $N_c=70$
        (second column), and $N_c=100$ (third column). The energy scale $Q$ at which which the correlations
    have been computed was chosen to be $Q=\gev{100}$.}
    \label{fig:corrmat}
\end{figure*}
An analogous way to verify that the compressed sets resulting from the GAN-enhanced methodology
reproduce more accurately the correlations of the prior PDF replicas is to compute the difference
in correlation matrices. That is, compute the correlation matrix for each set (prior, standard,
enhanced) and then compute the difference between the correlation matrix of the prior and the
standard (or enhanced respectively). Such studies are shown in~\Fig{fig:corrmat} where the
matrices are defined in a logarithmic $x$ grid with size $N_x=70$ points for each of the
$n_f=8$ light partons. The first row shows the difference between the correlation matrix
of the prior and the results from the standard compression, while the second row shows the
difference between the prior and the results from the generative-based compression. From the
first to the third row, we present results for the compressed set with size $N_c=50,70,100$.
As we go from left to right, we see that the correlation matrix is becoming lighter, indicating
an increase in similarity between the PDF correlations. This feature is seen on both compression
strategies. However, as we look from top to bottom, we can also see that the correlation matrices
in the bottom row are lighter than the ones on top. Although this is barely seen in the case
$N_c=50$, a minor difference can be seen at $N_c=70$ while the difference is clearly significant
for $N_c=100$. These confirm the results in~\Fig{fig:erfs}. Such a difference could be made more
apparent by projecting the values of the difference in correlation matrices into a histograms
and computing the mean and the standard deviation. \Fig{fig:projection} shows the histogram
projections of the difference in correlations given in~\Fig{fig:corrmat}.
\begin{figure*}[!h]
    \captionsetup[subfigure]{aboveskip=-1.5pt,belowskip=-1.5pt}
    \centering
    \begin{subfigure}{0.475\linewidth}
        \includegraphics[width=\linewidth]{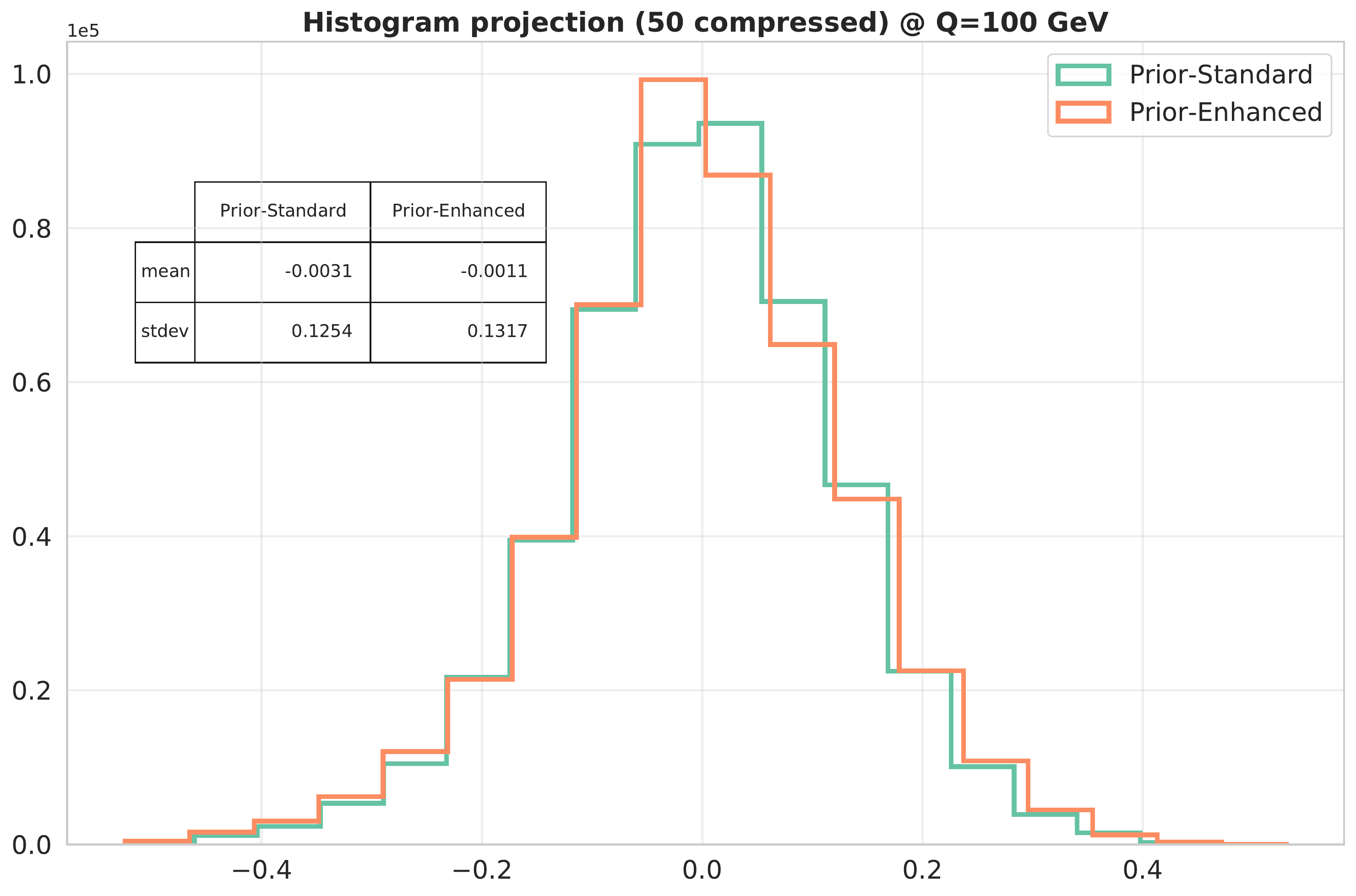}
    \end{subfigure}
    \begin{subfigure}{0.475\linewidth}
        \includegraphics[width=\linewidth]{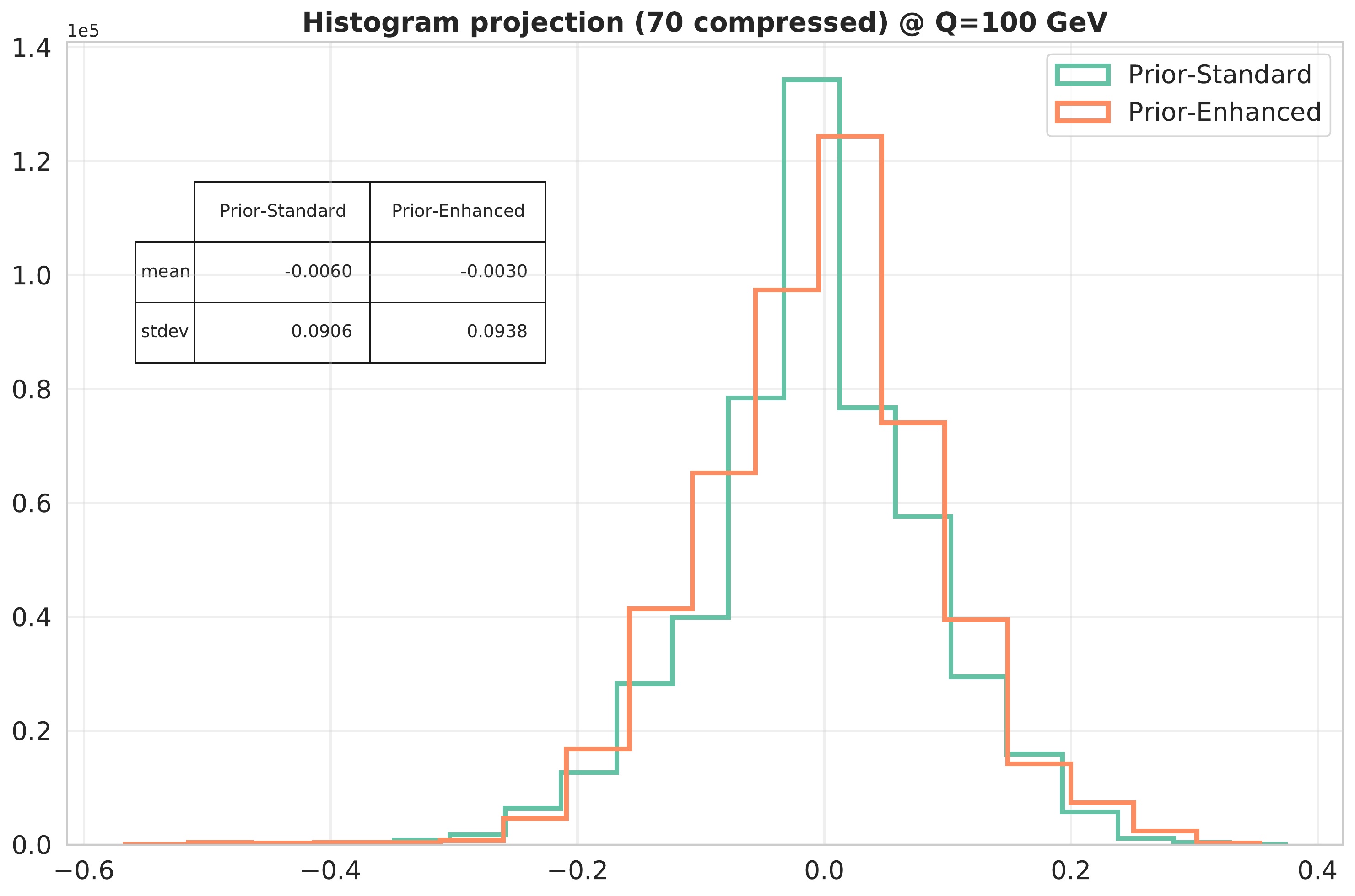}
    \end{subfigure}
    \begin{subfigure}{0.475\linewidth}
        \includegraphics[width=\linewidth]{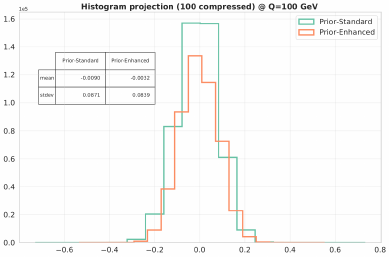}
    \end{subfigure}
    \caption{Histogram projection of the difference in correlation matrices given in~\Fig{fig:corrmat}.
        The green and orange lines represent the resulting differences for the standard and enhanced
        compression respectively, i.e the green lines represent the projection of the matrices in the first
    row of~\Fig{fig:corrmat} while the orange lines represent the projection of those in the second row.}
    \label{fig:projection}
\end{figure*}
The mean and standard
deviation values of the projected distributions are shown in the table on the top-left of the
plots. 

We see that the resulting means from the GAN-enhanced compression, for all sizes of the compressed sets,
are closer to zero. This confirms the previous results that GAN-enhanced compression yields
a more accurate representation of the correlations between the different partons.
So far, we only focused on the role that generative-based adversarial models play in the
improvement of the compression methodology. It was shown that, for all the statistical estimators
we considered (including lower and higher moments, various distance metrics, and correlations), the
new generative-based approach outperforms the standard compression. However, a question
still remains: Can the Generative Adversarial Neural Network truly replace partly (or fully) the fitting
procedure? In the next section, we try to shed some light on this question, which we hope
may lie the ground to future studies.

\section{Generalization capability of GANs for PDFs}
\label{sec:outlook}

Until this point, the usage of GAN has only been to reproduce, after a compression, the statistical
estimators of a large set of replicas as accurately as possible.
Instead, now we pose ourselves the following question:
can a GAN generate synthetic replicas beyond the finite size effects of the original ensemble?

The answer to this question could potentially open two new ways of using the present framework.
First, we can arbitrarily augment the density of the replicas of an existing large set in very little time.
Indeed, even with the newest NNPDF methodology~\cite{Carrazza_2019}, generating several thousand replicas can take
weeks of computational effort.
Second, once it is understood how finite-size effects affect synthetic replicas, one could, by generating as
many synthetic replicas as necessary, compress a PDF set down to a minimal set which is only limited by the target
accuracy even when the original set do not contain the appropriate discrete replicas
(as long as the relevant statistical information is present in the prior).

As mentioned in the previous sections, the convergence of the Monte Carlo PDF to the asymptotic result
depends strongly on the number of replicas (of the order of thousands).
Such a large number of replicas might be feasible in future releases of
NNPDF, based on the methodology presented in Ref.~\cite{Carrazza_2019}.
We limit ourselves here, however, to official NNPDF releases~\cite{Ball:2017nwa}.

We start by considering two disjoint sets of $N$ fitted replicas ($\mathrm{S}_1$ and $\mathrm{S}_2$),
and a set of synthetic replicas ($\mathrm{S}_3$) of the same size but determined
from GANs using a starting set of $N_0 < N$ fitted replicas.
Then, based on the various statistical estimators discussed previously,
we measure the distance between $\mathrm{S}_1$ and $\mathrm{S}_3$ and compare the result with the distance between
$\mathrm{S}_1$ and $\mathrm{S}_2$.
In order to estimate the uncertainty of the distance between two subsets of replicas we need to repeat the
exercise several times each time generating different subsets $\mathrm{S}_i$ (for $i=1,2,3$). This procedure is
clearly computationally intractable.
Instead, we address the problem using two resampling strategies:
the (delete-1) jackknifing~\cite{bradley1984use, 10.1093/sysbio/33.4.408, 10.1093/sysbio/34.4.404, 10.1093/auk/103.2.341}
and the logically similar (non-parametric) bootstrapping~\cite{efron1979, 566814, 10.2307/2345699}.
Both methods provide reliable estimates of the dispersion when the statistical model is not adequately known.

\begin{figure*}[!h]
    \captionsetup[subfigure]{aboveskip=-1.5pt,belowskip=-1.5pt}
    \centering
    \begin{subfigure}{0.495\linewidth}
        \includegraphics[width=\linewidth]{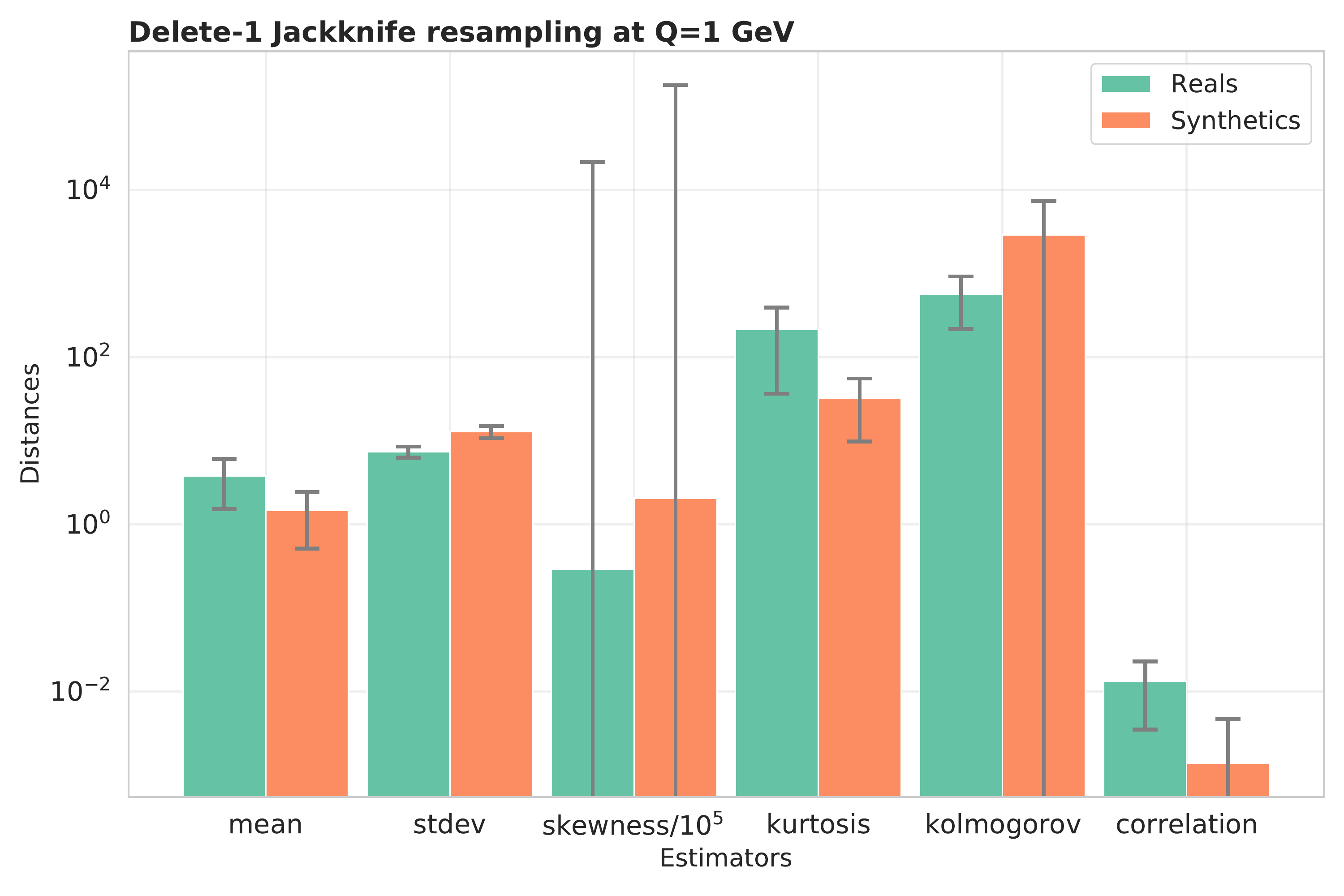}
    \end{subfigure}
    \hfil
    \begin{subfigure}{0.495\linewidth}
        \includegraphics[width=\linewidth]{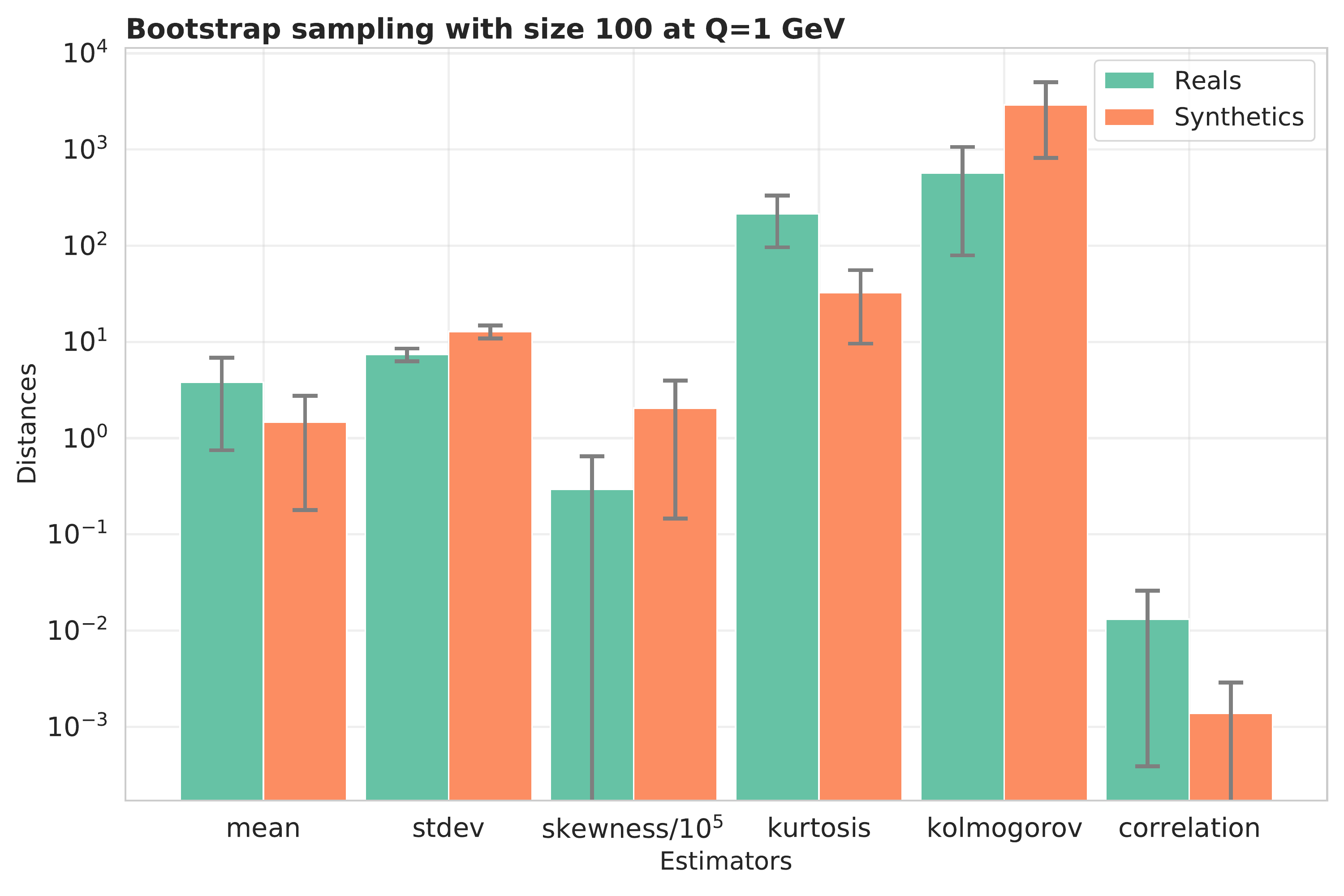}
    \end{subfigure}
    \caption{Comparison of the real and synthetic fits for various statistical estimators. The histograms
        are constructed by measuring the distance between $\mathrm{S}_1$-$\mathrm{S}_2$ (Reals) and
        $\mathrm{S}_1$-$\mathrm{S}_3$ (Synthetics). The error bars are computed by performing the delete-1 Jackknife
        (left) and the bootstrap resampling (right). In the bootstrap resampling, the results have been evaluated
    by performing $N_B=100$ bootstraps.}
    \label{fig:resampling}
\end{figure*}

In this study the three sets $\mathrm{S}_1$, $\mathrm{S}_2$ and $\mathrm{S}_3$ contain each $N\!=\!500$ replicas with
the difference that $\mathrm{S}_3$ was determined using $\texttt{ganpdfs}$ from a fit with $N_0=100$ replicas.
The results from both resampling methodologies are shown in~\Fig{fig:resampling}.
The histograms are constructed by measuring the distance between the two subsets of original replicas
($\mathrm{S}_1$-$\mathrm{S}_2$) and the distance between the synthetic replicas and one of original subsets ($\mathrm{S}_1$-$\mathrm{S}_3$).
The uncertainty bars are estimated using the Jackknife resampling and the bootstrap method.
From the Jackknife results, we observe that for most statistical estimators the synthetic and ``true'' fits
produce results which are compatible within uncertainties.
This is further confirmed by the results obtained through bootstrap resampling where
only the error bands for the standard deviation of the two sets do not overlap.

Given that inspecting~\Fig{fig:resampling} it is difficult to distinguish the two sets
$\mathrm{S}_2$ and $\mathrm{S}_3$ we hypothesize that this would also be the case for larger sets.
If true, the present framework could be used in any of the ways proposed at the beginning of the section.
We leave this for future work.

\section{Conclusions}
\label{sec:conclusions}

In this paper, we applied the generative adversarial strategies to the study of Parton
Distribution Functions. Specifically, we showed how such techniques could be used to
improve the efficiency of the compression methodology which is used to reduce the
number of replicas while preserving the same statistics. The implementation of the
GAN methodology required us to re-design the previously used compressor code into
a more efficient one.

It was shown that with the same size of compressed set, the GAN-enhanced compression methodology
achieves a more accurate representation of the underlying probability distribution than the standard compression.
This is due to the generation of synthetic replicas which increases the density of the
replicas for the compressor to choose from, reducing finite size effects.

Finally, in Section~\ref{sec:outlook} we entertain the idea of utilising these techniques
in order to generate larger PDF sets with reduced finite size effects.
We have compared real and synthetic Monte Carlo replicas using two different resampling techniques and
found that a GAN-enhanced set of replicas is statistically compatible with actual replicas.
However, further work is needed before synthetic PDFs can be used for precision studies. This basically
entails reproducing the same studies done here but with much larger samples in order to reduce statistical
fluctuations.

\section*{Acknowledgements}

The authors wish to thank Stefano Forte for innumerable discussions and suggestions
throughout the development of this project.
The authors thank Christopher Schwan, Zahari Kassabov and Juan Rojo
for careful readings of various iterations of the paper and useful comments.
This project is supported by the European Research Council under the European
Unions Horizon 2020 research and innovation Programme (grant agreement number 740006).

\appendix

\section{Benchmark of \texttt{pycompressor} against \texttt{compressor}}
\label{app:benchmark}

\begin{figure}[!h]
	\centering
	\includegraphics[width=\linewidth]{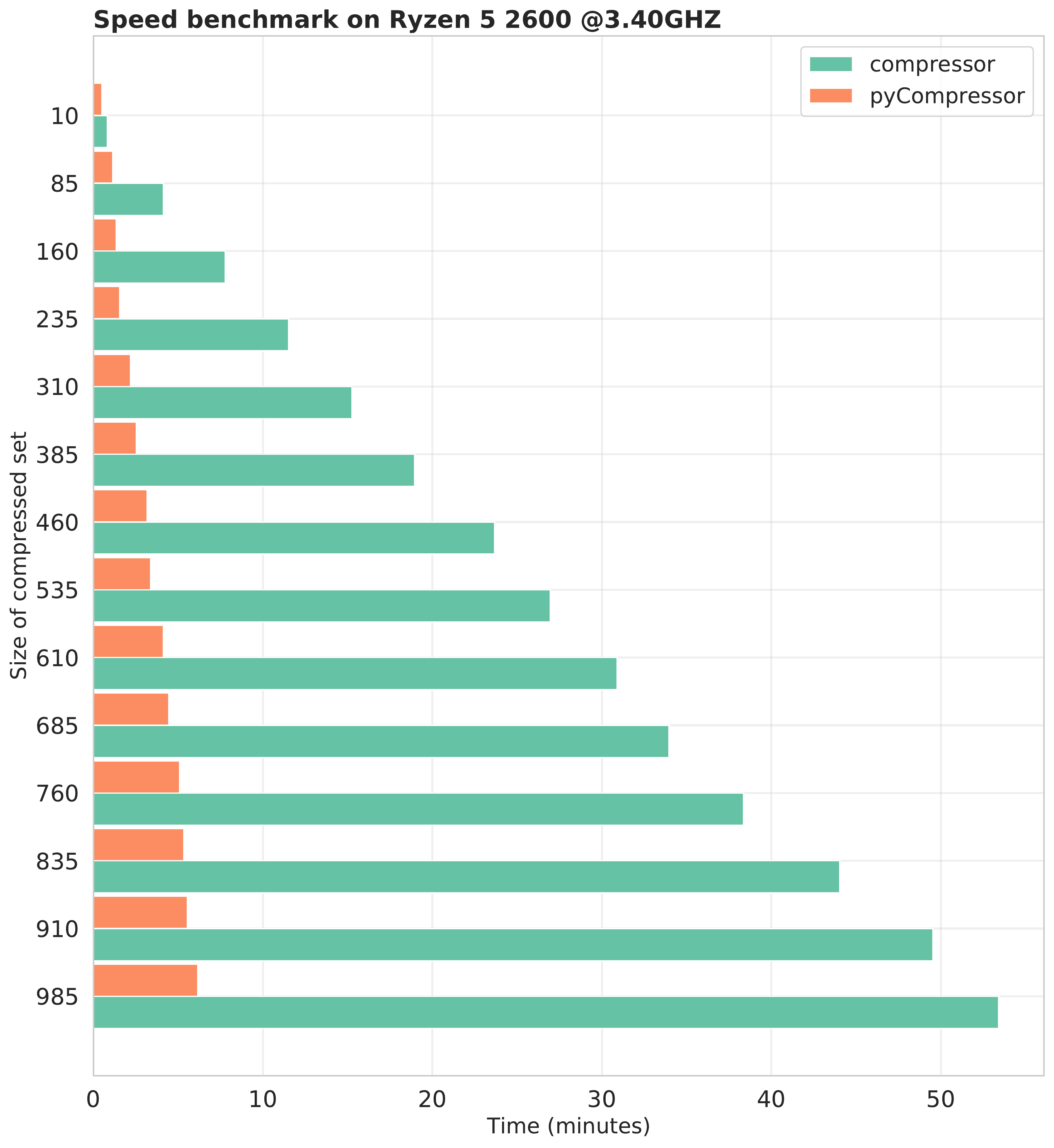}
	\caption{Speed benchmark comparing the \texttt{compresssor} and \texttt{pyCompressor} codes
		using the GA as a minimization for $1500$ iterations. For the purpose of this benchmarks, the
		parameters that enter into the GA are chosen to be exactly the same across both implementations.}
	\label{fig:compressor-benchmark}
\end{figure}

Among the benefits of the new code
is the gain in performance. For the purpose of quantifying this gain, we
compress a Monte Carlo PDF set with $N_p=1000$ replicas into sets with smaller sizes using
the same GA in the new and old codes.
The tests were run on a consumer-grade CPU (AMD Ryzen 5 2600 with 12 threads boosted
at 3.4 GHz) with 16 GB of memory. The results of the benchmark are shown in \Fig{fig:compressor-benchmark}.
The plots show the required time for both \texttt{compressor} and \texttt{pyCompressor} to
complete the same task. We see that the \texttt{pyCompressor} code is faster than
the previous implementation and that, as the size of the compressed set grows, the difference
in speed between the two implementations also increases.
This is mainly due to the fact that the new code automatically takes advantage of the
multicore capabilities of most modern computers.

On the flip side, parallelization comes at the expense of a slightly higher memory usage
(as shown in Table \ref{table:benchmark}), while not dramatic it can be a burden for very high number
of replicas.
\begin{table}[!h]
	\centering
	\begin{tabular}{c|cc}
		\hline
		& Nb. Threads & RAM usage (GB) \\
		\hline
		\texttt{compressor} & 1 & 1 \\
		\hline
		\texttt{pyCompressor} & 12 & 2.5 \\
		\hline
	\end{tabular}
	\vspace*{0.2cm}
	\caption{Comparison of average computing resources between ours and the previous implementation
		when compressing to $N_c=500$ replicas from a prior with $1000$ Monte Carlo PDF replicas.}
	\label{table:benchmark}
\end{table}

\section{Adiabatic Minimization}
\label{app:adiabatic}

\begin{figure}[!h]
    \centering
    \includegraphics[width=\linewidth]{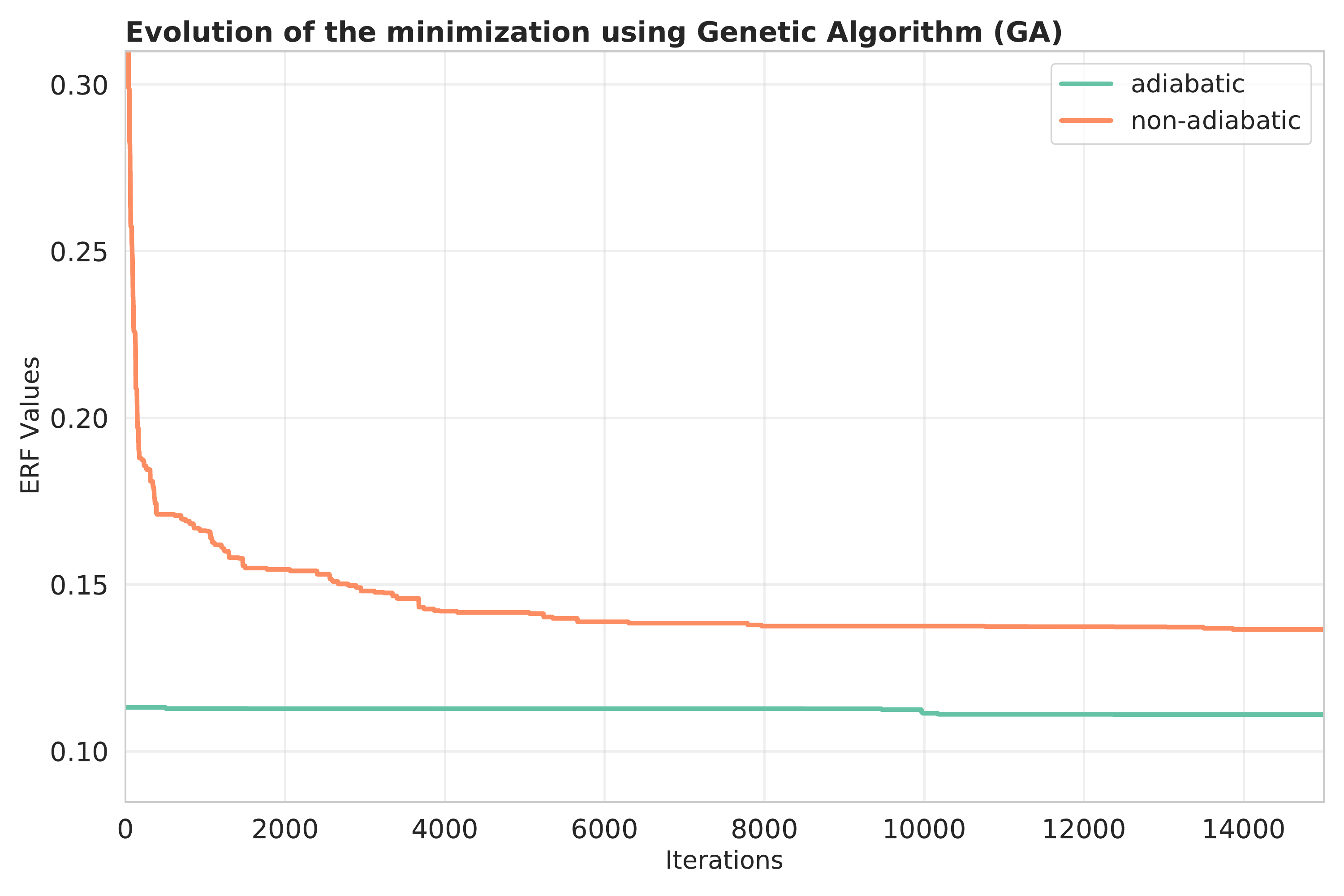}
    \caption{Comparison between adiabatic and standard minimization using the Genetic
        Algorithm (GA). The adiabatic minimization reaches stability faster while the standard
    minimization approach has not converged yet even at 15000 iterations.}
    \label{fig:adiabatic-minimization}
\end{figure}

Performing a compression from an enhanced set can be challenging due to the limitation
of the minimization algorithm. However, if results from the standard compression are
already provided, the compressor code provides a more efficient compression procedure
from the enhanced set with the means of an adiabatic minimization. The adiabatic
minimization for the enhanced compression consists on taking as a starting point the
space of replicas where the best from the standard compression was generated. Such a
minimization not only yields faster convergence (as shown in \Fig{fig:adiabatic-minimization}),
but also prevent the minimization algorithm to be trapped in some local minimum in case
the enhanced set contains statistical fluctuations.

\bibliographystyle{epj}
\bibliography{refs.bib}

\end{document}